\documentclass[final,times,onecolumn]{elsarticle}
\usepackage{graphicx}
\usepackage{amsmath}
\usepackage{amsbsy}
\usepackage{amsfonts}
\usepackage{amsthm}
\usepackage{amssymb}
\usepackage{color}

\begin{document}

\begin{frontmatter}
\title{Markov chain order estimation with parametric significance tests of conditional mutual information}
\author{Maria Papapetrou}\ead{mariapap@auth.gr}
\author{Dimitris Kugiumtzis}\ead{dkugiu@auth.gr}
\address{Department of Electrical and Computer Engineering, Aristotle University of Thessaloniki, 54124 Thessaloniki, Greece}

\begin{abstract}
Besides the different approaches suggested in the literature,
accurate estimation of the order of a Markov chain from a given
symbol sequence is an open issue, especially when the order is
moderately large. Here, parametric significance tests of
conditional mutual information (CMI) of increasing order $m$,
$I_c(m)$, on a symbol sequence are conducted for increasing orders
$m$ in order to estimate the true order $L$ of the underlying
Markov chain. CMI of order $m$ is the mutual information of two
variables in the Markov chain being $m$ time steps apart,
conditioning on the intermediate variables of the chain. The null
distribution of CMI is approximated with a normal and gamma
distribution deriving analytic expressions of their parameters,
and a gamma distribution deriving its parameters from the mean and
variance of the normal distribution. The accuracy of order
estimation is assessed with the three parametric tests, and the
parametric tests are compared to the randomization significance
test and other known order estimation criteria using Monte Carlo
simulations of Markov chains with different order $L$, length of
symbol sequence $N$ and number of symbols $K$. The parametric test
using the gamma distribution (with directly defined parameters) is
consistently better than the other two parametric tests and
matches well the performance of the randomization test. The tests
are applied to genes and intergenic regions of DNA sequences, and
the estimated orders are interpreted in view of the results from
the simulation study. The application shows the usefulness of the
parametric gamma test for long symbol sequences where the
randomization test becomes prohibitively slow to compute.
\end{abstract}

\end{frontmatter}

\section{Introduction}
\label{intro}
\par

Symbol sequences are directly observed on real-world processes,
such as DNA sequences and on-line transaction logs, but can also
be derived from discretization of time series. Sequence analysis,
initially developed mostly for biological applications
\cite{Durbin98}, has expanded with regard to both applications and
methodologies, and sequence mining techniques are constantly being
developed \cite{Dong07}. Here however, we concentrate on a
classical and fundamental problem that regards the memory of the
underlying mechanism to a symbol sequence. In the presence of
association in symbol sequences, the first step of the analysis is
to assume a Markov chain and estimate the order of the Markov
chain.

There are many Markov chain order estimators proposed and assessed in the literature. The Bayesian information criterion (BIC) and the Akaike
information criterion (AIC) are the two oldest and best known order estimators based on maximum likelihood \citep{Tong75,Katz81,Guttorp95}. Another
estimator is given by the maximal fluctuation method proposed by Peres-Shields \cite{Pere05} and modified by Dalevi and Dubhashi \cite{Dale05}, who
found that the Peres-Shields (PS) estimator is simpler, faster and more robust to noise than other criteria like AIC and BIC \cite{Dale05}. Other
order estimation schemes include the method of Men\'{e}ndez et al. \citep{Mene11}, which uses the $\phi$-divergence measures \cite{Pardo06}, the
method of global dependency level (GDL), also called relative entropy \cite{Baig11}, and the efficient determination criterion (EDC) \cite{Zhao01}.
Based on the information-related measures, and specifically the conditional mutual information (CMI), we recently proposed the order estimation by
means of randomization significance tests for CMI at increasing orders \cite{Papapetrou13}. In a somewhat similar way, Pethel et al. \cite{Pethel14}
propose a randomization test for the examined Markov chain order using the Chi-squared statistic.

In the approach of \cite{Papapetrou13} we made no assumption on the distribution of CMI. Here we propose the order estimation with parametric tests,
approximating the null distribution of CMI by normal and gamma distributions. We follow the bias correction and the approximation for the variance in
\cite{Miller55} and \cite{Roul99} and  approximate the distribution of mutual information with Gaussian distribution as an obvious possible choice
\citep{Paninski03,Hutt05}. We also consider the result in Goebel et al. \cite{Goeb05} that the statistic of mutual information (MI), and subsequently
CMI, follows gamma distribution. Finally, we consider a second gamma approximation with shape and scale parameter derived from the mean and variance
approximations of the normal distribution. We implement the three parametric significance tests for CMI and compare them to the randomization test of
\cite{Papapetrou13}, as well as other known Markov chain order estimation methods. Further, we attempt to assess the Markov chain order of DNA
sequences and infer for short and long range correlation on the basis of the parametric and randomization CMI testing. A systematic investigation of
long range correlation of DNA sequences using the CMI approach is reported in \cite{Papapetrou14}.

The structure of the paper is as follows. First, in Section~\ref{sec:CMI}, CMI is defined and estimated on symbol sequences. Parametric significance
tests for CMI of increasing orders are presented, approximating the null distribution of CMI by the normal and gamma distributions. In
Section~\ref{sec:Simulations}, we assess the efficiency of the parametric tests in estimating the Markov chain orders and compare them to other known
methods. In Section~\ref{sec:DNA}, we apply the parametric and randomization tests to DNA sequences, and in Section~\ref{sec:Conclusions}, the
results are discussed and the main conclusions are drawn.

\section{Conditional Mutual Information and Markov Chain Order Estimation}
\label{sec:CMI}

We start with the definition of entropy, mutual information (MI)
and conditional mutual information (CMI) for Markov chains. Let
$\{x_t\}$ denote a symbol sequence generated by a Markov chain
$\{X_t\}$, $t\geq 1$, of an unknown order $L \geq 1$ in a discrete
space of $K$ possible states $A=\{a_1,\ldots,a_K\}$, $p(x_t)$ the
probability of $x_t \in A$ occurring in the chain,
$\mathbf{X}_t=[X_t,X_{t-1},\ldots,X_{t-m+1}]$ a vector (word) of
$m$ successive variables of the Markov chain and $p(\mathbf{x}_t)$
the probability of a word
$\mathbf{x}_t=\{x_t,x_{t-1},\ldots,x_{t-m+1}\} \in A^{m}$
occurring in the chain. The entropy of a random variable of the
Markov chain $X_t$ is $H(X_t) = - \sum_{x_t} p(x_t)\ln{p(x_t)}$
and the entropy of a word $\mathbf{X}_t$ is $H(\mathbf{X}_t) = -
\sum_{x_t,\ldots,x_{t-m+1}} p(\mathbf{x}_t)\ln p(\mathbf{x}_t)$.
The MI of two random variables in the Markov chain being $m$ time
steps apart is \cite{Cover91}
\begin{align}
I(m) & = I(X_t;X_{t-m}) = H(X_t) + H(X_{t-m}) - H(X_t,X_{t-m}) \notag \\
& = \sum_{x_t,x_{t-m}} p(x_t,x_{t-m}) \ln{ \frac{p(x_t,x_{t-m})}{p(x_t)p(x_{t-m})}}\notag
\end{align}
and quantifies the amount of information for the one variable given the other variable.

The fundamental property of a Markov chain of order $L$ is
\begin{equation}
p(X_t|X_{t-1},X_{t-2},\ldots,X_{t-L},X_{t-L-1},\ldots) = p(X_t|X_{t-1},X_{t-2},\ldots,X_{t-L}),
\label{eq:Markovproperty}
\end{equation}
meaning that the distribution of the variable $X_t$ of the Markov chain at time $t$ is determined only in terms of the preceding $L$ variables of the chain. It is noted that $I(m)$ for $m>L$ may not drop to zero due to the existence of MI between the intermediate variables. Thus for estimating $L$ we consider CMI that accounts for the intermediate variables. CMI of order $m$ is defined as the mutual information of $X_t$ and $X_{t-m}$ conditioning on $X_{t-m+1},\ldots,X_{t-1}$ \cite{Cover91}
\begin{align}
I_c(m) & = I(X_t;X_{t-m}|X_{t-1},\ldots,X_{t-m+1}) \notag \\
& = -H(X_t,\ldots,X_{t-m})+H(X_{t-1},\ldots,X_{t-m}) + H(X_t,\ldots,X_{t-m+1}) -H(X_{t-1},\ldots,X_{t-m+1}) \notag  \\
& = \sum_{x_t,\ldots,x_{t-m}} p(x_t,\ldots,x_{t-m}) \ln{ \frac{p(x_t|x_{t-1},\ldots,x_{t-m})}{p(x_t|x_{t-1},\ldots,x_{t-m+1})}}. \label{eq:CMI}
\end{align}
CMI coincides with MI for successive random variables in the chain, $I_c(1)=I(1)$.

From the Markov chain property in (\ref{eq:Markovproperty}), for $m > L$ the logarithmic term in the sum of (\ref{eq:CMI}) is zero and thus $I_c(m) =
0$. On the other hand, for $m \le L$, we expect in general the two variables $m$ time steps apart be dependent given the $m-1$ intermediate
variables, and $I_c(m) > 0$. It is possible that $I_c(m) = 0$ for $m < L$, but not for $m=L$, as then the Markov chain order would not be $L$.  So,
increasing the order $m$, we expect in general when $I_c(m) > 0$ and $I_c(m+1) = 0$ to have $m=L$. To account for complicated and rather unusual
cases where $I_c(m+1) = 0$ occurs for $m+1<L$, we can extend the condition $I_c(m) > 0$ and $I_c(m+1) = 0$ to require also $I_c(m+2) = 0$, and even
further up to some maximum order.

\subsection{Parametric tests for Markov chain order estimation}

The estimate of entropy, MI and CMI from a symbol sequence $\{x_t\}_{t=1}^N$ of length $N$ is derived directly by the maximum likelihood estimate of
the probabilities given simply by the relative frequencies. As entropy and MI are fundamental quantities of information theory with many
applications, there is rich literature about the statistical properties and distribution of their estimates. Some works have focused on correcting
the bias in the estimation of entropy \citep{Miller55,Grass88,Schm93,Roul99,Schu02,Paninski03,Bonachela08,Lesne09}, whereas other
works give approximations with parametric distributions \citep{Pard95,Wolp95,Roul99,Goeb05,Hutt05}. For example, Roulston \cite{Roul99} estimates the
bias and variance of the observed entropy and gives evidence for normal distribution of the estimates. Pardo \cite{Pard95} shows that, under
different assumptions, the MI estimate is either normal or a linear combination of $\chi^2$ variables, while Goebel et al. \cite{Goeb05} using a
second-order Taylor approximation of the MI estimate derives a gamma distribution, and the same does for CMI. Hutter and Zaffalon \cite{Hutt05} use a
Bayesian framework with Dirichlet prior distribution to obtain the posterior distribution of MI estimate and derive expressions for its mean and
variance.

For simplicity in the expressions below, we assign $X$ for $X_t$,
$Y$ for $X_{t-m}$ and $Z$ for the vector variable of
$X_{t-1},\ldots,X_{t-m+1}$. The number of observed distinct
symbols of $X$ and $Y$ are $K$ but for $Z$ there may be less
observed distinct words than $K^{m-1}$ denoted $K_Z$. Similarly,
$K_{XZ}$, $K_{YZ}$ and $K_{XYZ}$ denote the number of observed
distinct words concatenating the respective indexed variables.
Note that the words $XZ$ and $YZ$ correspond to
$X_t,X_{t-1},\ldots,X_{t-m+1}$ and
$X_{t-1},\ldots,X_{t-m+1},X_{t-m}$, respectively, and therefore we
have $K_{XZ}=K_{YZ}$ (discrepancy by one may occur due to edge
effect). Further, we denote $N_m=N-m$.

\subsubsection{Approximation with normal distribution (ND)}
An expression for the mean of the entropy estimate $\hat{H}(X)$, $\langle\hat{H}(X) \rangle$, is given by the bias correction of Miller \cite{Miller55}
\begin{equation}
\langle\hat{H}(X) \rangle= H(X) - \frac{K-1}{2N}.
\label{eq:Hestimate}
\end{equation}
The same expression holds for vector variables (words) adjusting
accordingly the number of observed distinct words.
The mean of the CMI estimate $\hat{I}_c(m)$ can thus be derived by
substituting the mean of entropy estimates of (\ref{eq:Hestimate})
in the expression of CMI of (\ref{eq:CMI})
\begin{equation}
\langle \hat{I}_c(m) \rangle = I_c(m) + \frac{K_{XYZ} -K_{ZX}-K_{YZ}+ K_Z}{2N_m}.
\label{eq:meanCMI}
\end{equation}

For the CMI variance, we follow the variance approximation for MI
in \cite{Roul99}. We start with the error propagation formula for
$\hat{I}_c(m)$
\begin{equation}
V[\hat{I}_c(m)]= \sum_{u=1}^{K}\sum_{v=1}^{K}\sum_{w=1}^{K_Z}\left(\frac{\partial \hat{I}_c(m)}{\partial n_{uvw}}\right)^2V[n_{uvw}],
\label{eq:Var}
\end{equation}
where $V[\centerdot]$ denotes the variance and $n_{uvw}$ is the frequency of the concatenated word of $XYZ$ that corresponds to the indices $uvw$. We
want to express $\hat{I}_c(m)$ in (\ref{eq:Var}) in terms of the observed probabilities (relative frequencies) of joints words of $XYZ$,
$q_{ijk}=n_{ijk}/N_m$, and the marginal probabilities, e.g. $q_{\cdot{jk}}=\sum_{i=1}^{K}q_{ijk}$ and
$q_{\cdot{\cdot{k}}}=\sum_{i=1}^{K}\sum_{j=1}^{K}q_{ijk}$. Substituting these probabilities in the four entropy terms in (\ref{eq:CMI}) we get
\begin{equation}
\hat{I}_c(m) = \sum_{i=1}^{K}\sum_{j=1}^{K}\sum_{k=1}^{K_Z}q_{ijk}\ln q_{ijk} -\sum_{i=1}^{K}\sum_{k=1}^{K_Z}q_{i\cdot{k}}\ln q_{i\cdot{k}} -
\sum_{j=1}^{K}\sum_{k=1}^{K_Z}q_{\cdot{jk}}\ln q_{\cdot{jk}} + \sum_{k=1}^{K_Z}q_{\cdot{\cdot{k}}} \ln q_{\cdot{\cdot{k}}}. \label{eq:CMIq}
\end{equation}
Differentiation of the observed joint and marginal probabilities
in (\ref{eq:CMIq}) with respect to $n_{uvw}$ gives the following
expressions
\begin{align}
\frac{\partial q_{ijk}} {\partial n_{uvw}} & =-\frac{n_{ijk}}{N_m^2}+\frac{\delta_{iu}\delta_{jv}\delta_{kw}}{N_m}, \label{eq:difa}
\end{align}
\begin{align}
\frac{\partial q_{\cdot{jk}}}{\partial n_{uvw}} & =-\frac{1}{N_m^2}\sum_{i=1}^{K}n_{ijk}+\frac{\delta_{jv}\delta_{kw}}{N_m}, \label{eq:difb}
\end{align}
\begin{align}
\frac{\partial q_{i\cdot{k}}}{\partial n_{uvw}} &  =-\frac{1}{N_m^2}\sum_{j=1}^{K}n_{ijk}+\frac{\delta_{iu}\delta_{kw}}{N_m}, \label{eq:difc}
\end{align}
\begin{align}
\frac{\partial q_{\cdot{\cdot{k}}}}{\partial n_{uvw}} & = -\frac{1}{N_m^2} \sum_{i=1}^{K}\sum_{j=1}^{K}n_{ijk}+\frac{\delta_{kw}}{N_m},
\label{eq:difd}
\end{align}
where $\delta_{mn}$ is the Kronecker delta defined as $\delta_{mn}=1$ when $m=n$ and $\delta_{mn}=0$ when $m \neq n$. Substitution of
(\ref{eq:difa}-\ref{eq:difd}) into (\ref{eq:Var}) gives
\begin{equation}
V[\hat{I}_c(m)] = \sum_{u=1}^{K}\sum_{v=1}^{K}\sum_{w=1}^{K_Z}\frac{1}{N_m^2}\left(-\ln q_{uvw} + \ln q_{u\cdot{w}} + \ln q_{\cdot{vw}} - \ln
q_{\cdot{\cdot{w}}}+\hat{I}_c\right)^2 V[n_{uvw}]. \notag
\end{equation}
The observed frequency $n_{uvw}$ of the concatenated word of $XYZ$
is itself a binomial random variable, $n_{uvw} \sim
B(N_m,q_{uvw})$, considering the occurrence of the word as success
with probability $q_{uvw}$ and as number of trials the number
$N_m$ of possible words of length $m$ in the symbol sequence. Thus
the variance of $n_{uvw}$ is $V[n_{uvw}] = N_m q_{uvw}(1-q_{uvw})$
and substituting it in the expression above we have the final
expression of the variance of $\hat{I}_c$
\begin{equation}
V[\hat{I}_c(m)] = \sum_{u=1}^{K}\sum_{v=1}^{K}\sum_{w=1}^{K_Z}\frac{1}{N_m}\left(-\ln q_{uvw} + \ln q_{u\cdot{w}} + \ln q_{\cdot{vw}} - \ln
q_{\cdot{\cdot{w}}}+\hat{I}_c\right)^2 q_{uvw}(1-q_{uvw}). \label{eq:VarCMI}
\end{equation}
Thus $V[\hat{I}_c(m)]$ is directly derived when the observed
probabilities $q_{uvw}$ are first computed on the symbol sequence.

In \cite{Roul99}, similar expressions for the mean and variance of
$I(m)$ were derived to define the normal approximation of the MI
distribution. Similarly, we assume that the distribution of CMI
follows approximately the normal distribution (denoted hereafter
ND)
\begin{equation}
\hat{I}_c(m)\sim N_m(\langle \hat{I}_c(m) \rangle,V[\hat{I}_c(m)]),
\label{eq:NormDistr}
\end{equation}
where $\langle \hat{I}_c(m) \rangle$ is given by
(\ref{eq:meanCMI}) and $V[\hat{I}_c(m)]$ by (\ref{eq:VarCMI}).

\subsubsection{Approximation with gamma distribution (GD1)}

Goebel et al. \cite{Goeb05} approximate the expression of
distribution for CMI by means of a second order Taylor series
expansion. The second order Taylor approximation of MI about
$p(x,y)\equiv p(x)p(y)$ (assuming independence) is
\[
I(X,Y)=\frac{1}{2\ln2}\mathop{\sum_{x\in X}\sum_{y\in Y}}\frac{(p(x,y)-p(x)p(y))^2}{p(x)p(y)},
\]
and the estimate $\hat{I}(X,Y)$ is defined accordingly substituting $p(x,y)$ with the observed probability $q_{ij}=n_{ij}/N$, where $i,j\in \{1,\ldots,K\}$, and the same for the marginal probabilities.
The expression for MI is related to the $\chi^2$ statistic of the standard chi-square test of independence, which is defined as
\[
\chi^2 =\mathop{\sum_{i=1}^{K}\sum_{j=1}^{K}}\frac{(n_{ij}-(n_{\cdot{j}}n_{i{\cdot}})/N)^2}{(n_{\cdot{j}}n_{i{\cdot}})/N},
\]
and follows a chi-square distribution with $(K-1)(K-1)$ degrees
of freedom under the assumption of independence of $X$ and $Y$.
The above equations are related by $\chi^2=2N\ln2\hat{I}(X,Y)$,
from which the approximate gamma distribution
$\Gamma\left((K-1)^2/2,1/(N\ln2)\right)$ of $\hat{I}(X,Y)$ is
established \cite{Goeb05}. Further, it follows that
$\hat{I}(X,Y|Z)$ is approximately gamma distributed (denoted
hereafter GD1)
\begin{equation}
\hat{I}_c(m)=\hat{I}(X,Y|Z) \sim \Gamma\left(\frac{K_Z}{2}(K-1)(K-1),\frac{1}{N\ln2}\right). \label{eq:Gamma1}
\end{equation}

\subsubsection{Approximation with gamma distribution and moments from normal distribution
(GD2)}

It is known that a gamma distribution $\Gamma(\alpha,\beta)$ with shape parameter $\alpha$ being a positive integer and scale parameter $\beta$, can
be approximated by a normal distribution $N(\alpha\beta,\alpha\beta^2)$ if $\alpha$ is sufficiently large \citep{Johnson94, Bill95}. Reversing this
result, using the mean and variance of $\hat{I}(X,Y|Z)$ in (\ref{eq:meanCMI}) and (\ref{eq:VarCMI}), respectively, we can estimate the parameters of
gamma distribution and obtain approximately the gamma distribution for $\hat{I}(X,Y|Z)$ (denoted hereafter GD2)
\begin{equation}
\hat{I}_c(m)=\hat{I}(X,Y|Z)\sim \Gamma\left(\frac{\hat{I}_c^2}{V[\hat{I}_c]},\frac{V[\hat{I}_c]}{\hat{I}_c}\right). \label{eq:Gamma2}
\end{equation}

\subsubsection{Parametric tests for the significance of CMI}

Having determined the three parametric approximations for the
distribution of $\hat{I}_c(m)$, we use them as null distributions
for the null hypothesis H$_0$: $I_c(m)=0$. Given that it
always holds $I_c(m) \ge 0$, all three parametric tests are
one-sided. We compute the $p$-value from the cumulative function
of the null distributions ND, GD1 and GD2 of the observed CMI
$\hat{I}_c(m)$, and we reject H$_0$ if the $p$-value is less than
the nominal significance level $\alpha$ (we set $\alpha=0.05$ for
all simulations below). We apply the significance test for
increasing orders $m$ until we obtain rejection of H$_0$ for $m$ and
no rejection of H$_0$ for $m+1$, and then the estimate of $L$ is $m$.
The parametric tests
are denoted as ND, GD1 and GD2 corresponding to the respective
null distributions.

\subsection{Randomization test for the significance of CMI(RD)}

In a recent work \cite{Papapetrou13}, we developed a randomization
significance test for $I_c(m)=0$ and formed the null distribution
for H$_0$: $I_c(m)=0$, empirically. For the randomization test, we
first generate $M$ randomized symbol sequences
$\{x_t^{*1}\}_{t=1}^N,\ldots, \{x_t^{*M}\}_{t=1}^N$ by random
permutation of the initial sequence $\{x_t\}_{t=1}^N$. Then we
compute $\hat{I}_c(m)$ on the original symbol sequence, denoted
$\hat{I}_c^0(m)$, and on the $M$ randomized sequences, denoted
$\hat{I}_c^{*1}(m),\ldots,\hat{I}_c^{*M}(m)$. Finally, we reject
$\mbox{H}_0$ if $\hat{I}_c^0(m)$ is at the right tail of the
empirical null distribution formed by
$\hat{I}_c^{*1}(m),\ldots,\hat{I}_c^{*M}(m)$. To assess this we
use rank ordering, where $r^0$ is the rank of $\hat{I}_c^0(m)$ in
the ordered list of the $M+1$ values, assuming ascending order.
The $p$-value of the one-sided test is $1 -
(r^0-0.326)/(M+1+0.348)$ using the correction in \cite{Yu01}. The
randomization test is denoted as RD.

\subsection{Parametric and randomization significance test for CMI}

\begin{figure}[htb]
\centerline{\hbox{\includegraphics[width=6cm]{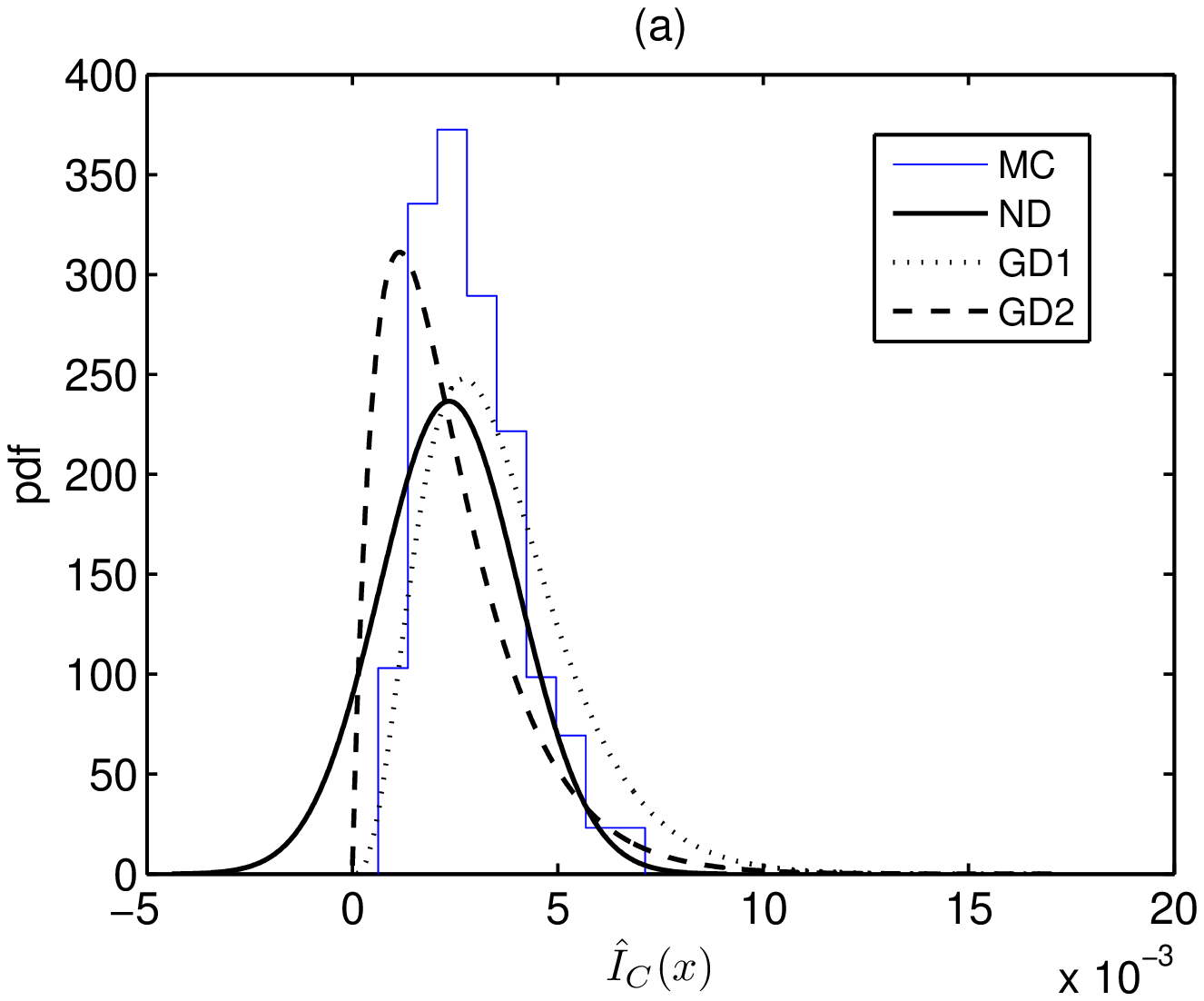}
\includegraphics[width=6cm]{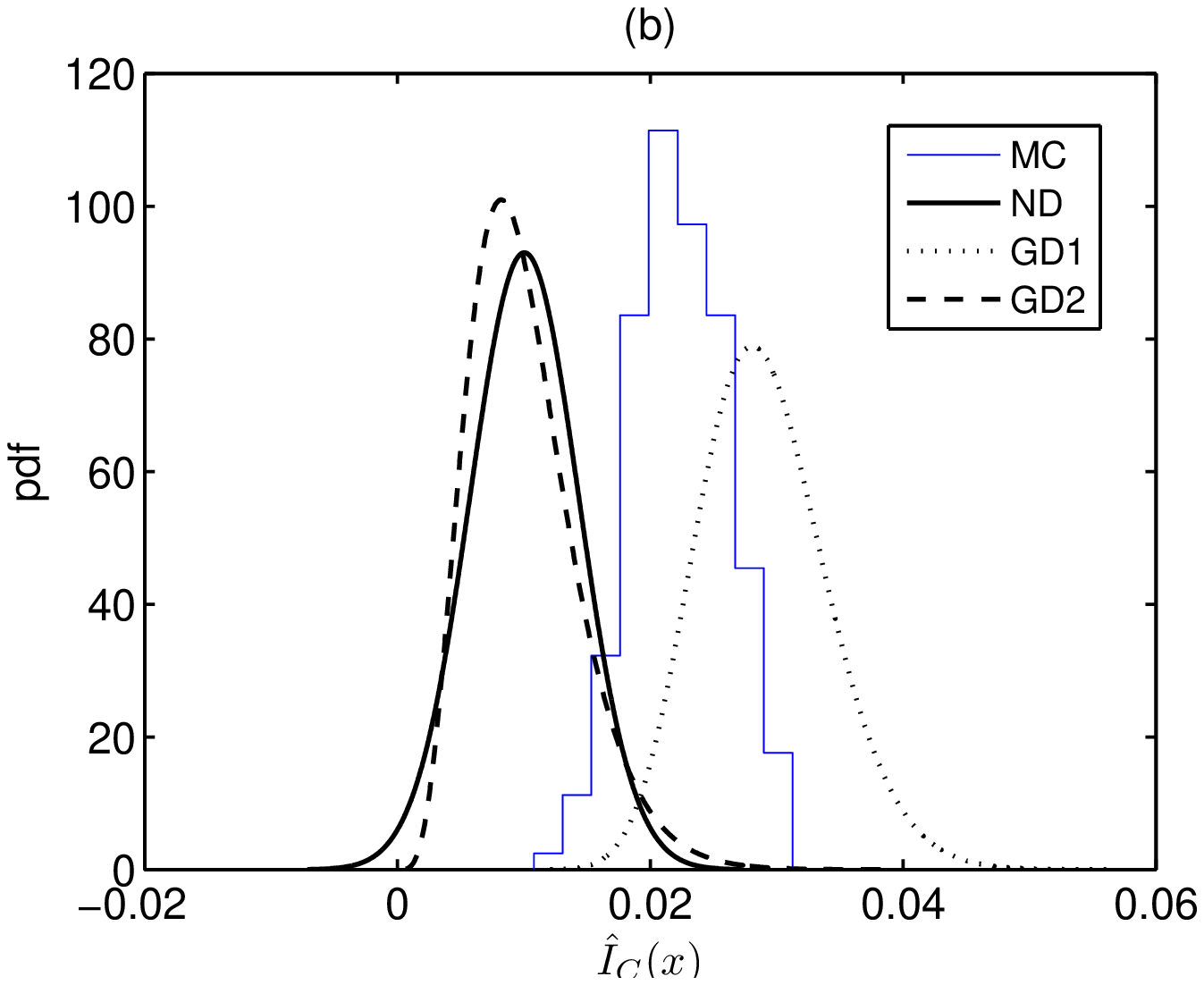}}}
\centerline{\hbox{\includegraphics[width=6cm]{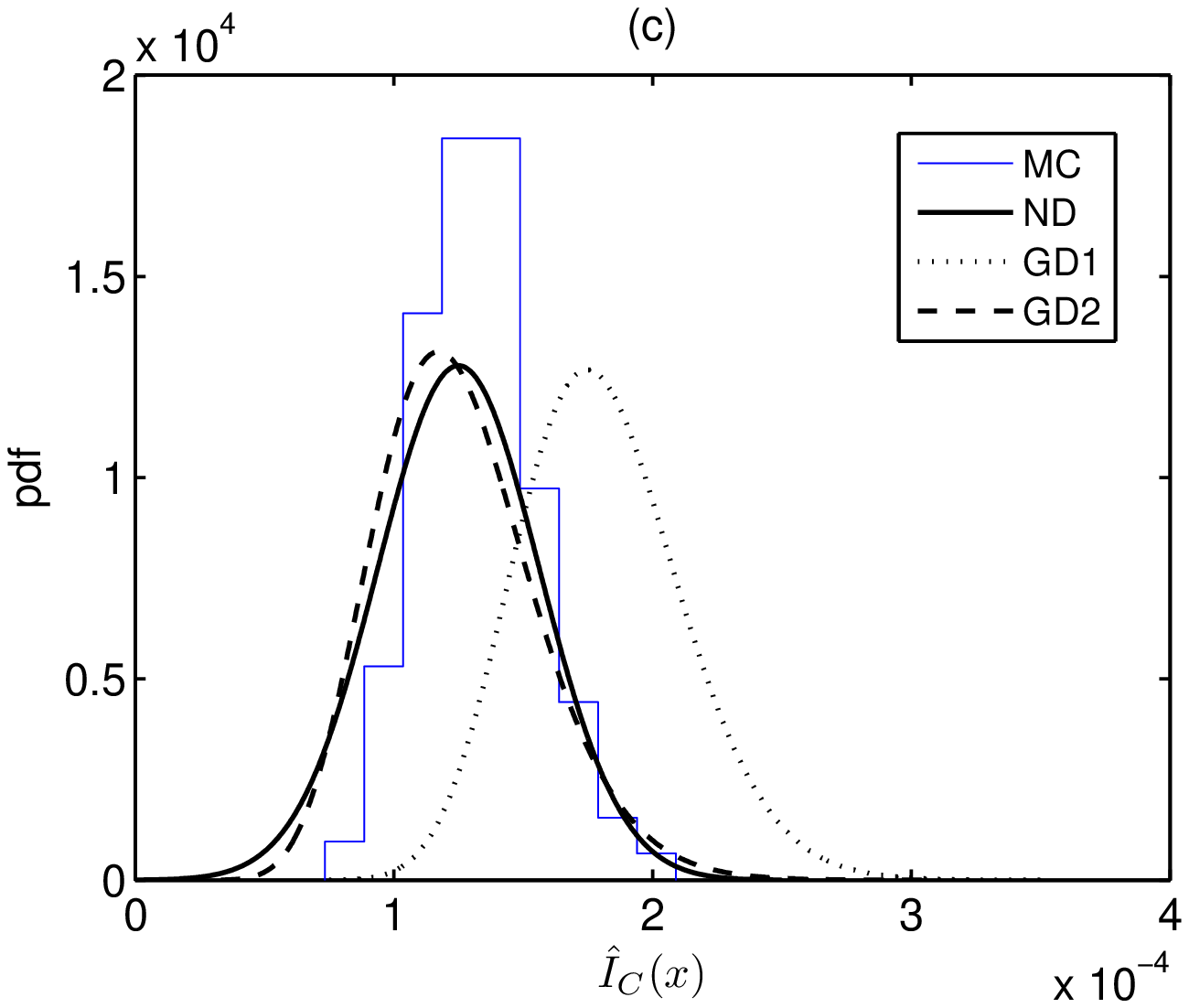}}} \caption{The true distribution of $\hat{I}_c(L)$ and the three approximations ND, GD1 and
GD2, formed from 1000 Monte Carlo realizations of a Markov chain of $K=2$. (a) $L=3$ and $N=1600$, (b) $L=6$ and $N=1600$, and (c) $L=6$ and
$N=256000$.} \label{fig:K2N1600MC1000RandomMatrix}
\end{figure}

Here, we show the differences of the distributions ND, GD1 and GD2
in approximating CMI with an example of two Markov chains of order
$L=3$ and $L=6$, number of symbols $K=2$ and symbol sequence
length $N=1600$ and $N=256000$. The true distribution of CMI,
$\hat{I}_c(m)$, for order $m=L$, is approximated by 1000 Monte
Carlo realizations, as shown in
Figure~\ref{fig:K2N1600MC1000RandomMatrix} \textcolor{red}{with
the broken line displaying the histogram}. The three approximating
distributions are drawn setting their parameters as defined in
(\ref{eq:NormDistr}), (\ref{eq:Gamma1}) and (\ref{eq:Gamma2}) to
the corresponding average values from the 1000 realizations. As
shown in Figure~\ref{fig:K2N1600MC1000RandomMatrix}, all three
approximations match quite well the true distribution of CMI for
$L=3$ (see Figure~\ref{fig:K2N1600MC1000RandomMatrix}a), but for
$L=6$ ND and GD2 lie to the left while GD1 tends to lie to the
right of the true distribution (see
Figure~\ref{fig:K2N1600MC1000RandomMatrix}b). It seems that as the
chain order increases\textcolor{red}{, the approximations of} ND,
GD1 and GD2 tend to deviate more from the true distribution. The
match tends to be regained by increasing the chain length. Indeed
when we increase the sequence length to $N=256000$, all
distributions translate closer to zero and have smaller width, as
expected, and the distributions of ND and GD2 approximate better
the true distribution, whereas the distribution of GD1 is still at
the left of the true distribution (see
Figure~\ref{fig:K2N1600MC1000RandomMatrix}c). The latter indicates
that the significance test with GD1 is more conservative, and for
this case the probability of rejection of H$_0$ is expected to be
smaller than the nominal significance level.

\begin{figure}[htb]
\centerline{\hbox{\includegraphics[width=6cm]{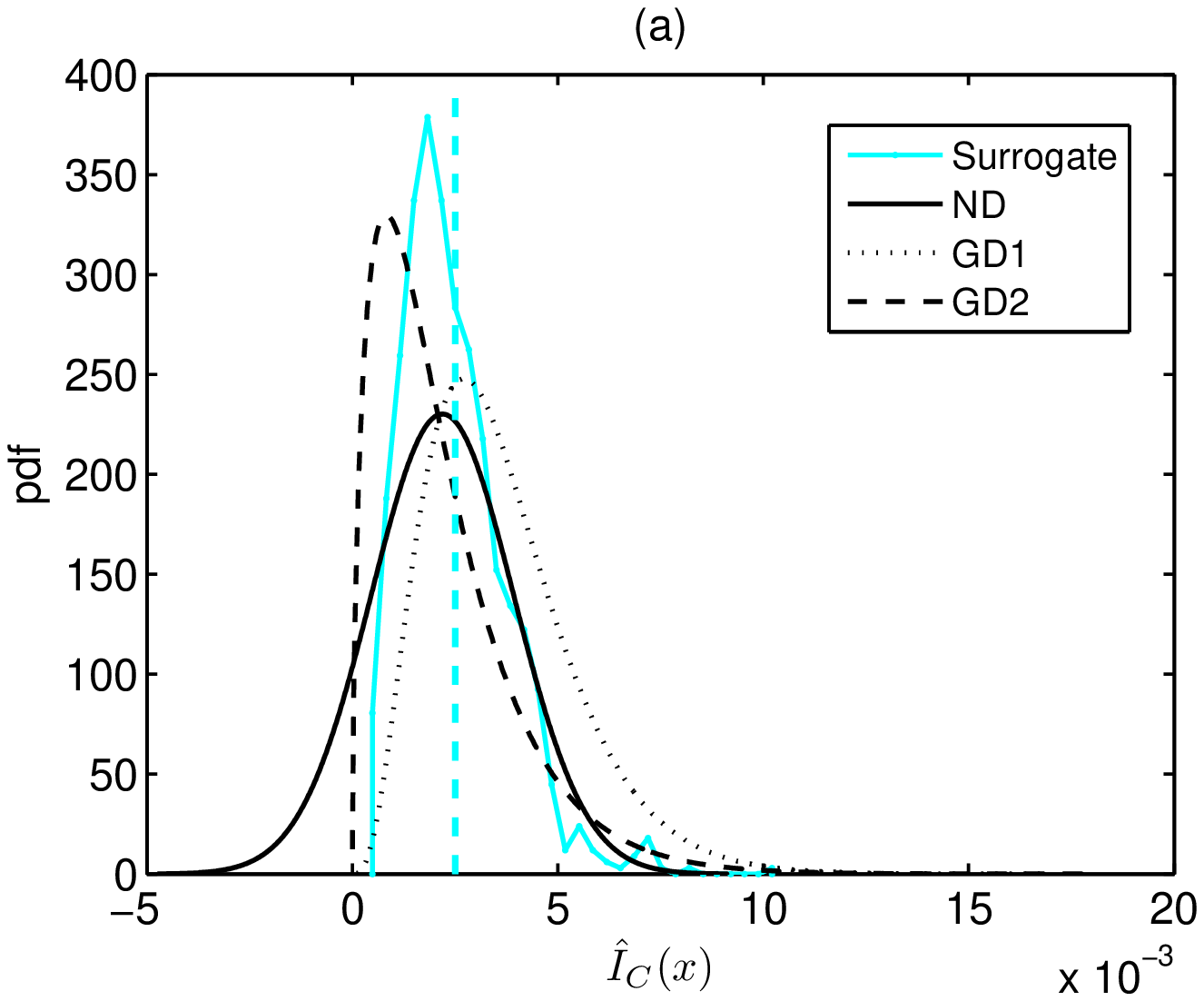}
\includegraphics[width=6cm]{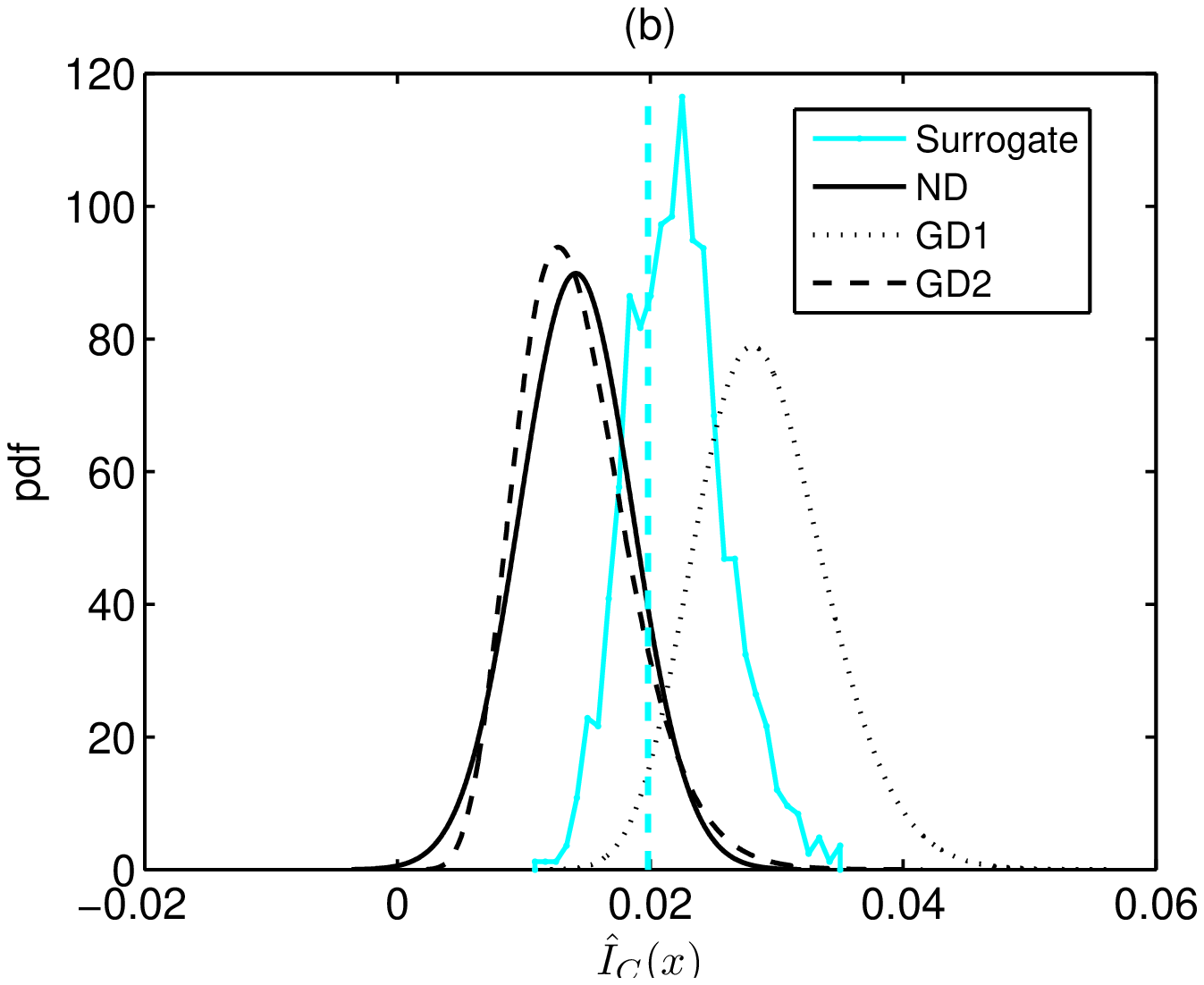}}}
\caption{The three parametric approximations of the null distribution of $\hat{I}_c(L)$ and the distribution formed by $M=1000$ surrogates for one
realization of length $N=1600$, $K=2$, and (a) $L=3$, (b) $L=6$. The observed value of $\hat{I}_c(L)$ is shown by a vertical dashed line.}
\label{fig:K2N1600RandomMatrix}
\end{figure}

The three parametric tests are then compared to the randomization test. For one realization of the same Markov chains with $L=3$ and $L=6$ ($N=1600$), the three parametric null distributions and the null distribution formed by CMI values from 1000 surrogates are shown in Figure~\ref{fig:K2N1600RandomMatrix}.
For $L=3$ in Figure~\ref{fig:K2N1600RandomMatrix}a, the H$_0$ of $I_c(L)=0$ is not rejected for any of the one-sided tests with the statistic
$\hat{I}_c$ because all four distributions cover well the observed value of $\hat{I}_c(L)$ (shown by a vertical dashed line in
Figure~\ref{fig:K2N1600RandomMatrix}). On the contrary, for $L=6$, $\hat{I}_c(L)$ lies towards the right tail of ND and GD2 distribution tending to
give false rejection, and on the left of the RD and GD1 distributions giving correctly no rejection (see Figure~\ref{fig:K2N1600RandomMatrix}b).
Moreover, the null distribution of GD1 is further to the right of the observed value $\hat{I}_c(L)$ than the null distribution of RD, suggesting that
the test with GD1 may be more conservative than with RD for this setting.

\section{Monte Carlo Simulations}
\label{sec:Simulations}

We evaluate the three parametric tests (ND, GD1 and GD2) and the randomization test (RD) using Monte Carlo simulations for varying Markov chain order
$L$, number of symbols $K$ and symbol sequence length $N$. We also compare the RD and parametric tests with four known criteria for the estimation of
$L$: the Akaike's information criterion (AIC) \citep{Tong75,Guttorp95}, the Bayesian information criterion (BIC)
\citep{Guttorp95}, the criterion of Dalevi and Dubashi which is based on the Peres and Shields estimator (PS)
\citep{Pere05,Dale05} and the criterion of Men\'{e}ndez et al. (Sf) \citep{Mene06,Mene11}. For each parameter setting, we use $100$ realizations, and $M=1000$ randomized sequences for each realization for the randomization test. The Markov chain order is sought in the range $m=1,\ldots,L+1$ by applying each of the four significance tests of $I_c(m)$ for increasing order $m$, as well as the aforementioned criteria. In the first simulation setup, Markov chains are derived by randomly selected transition probability matrices of given order $L$, while in the second simulation setup,
Markov chains are derived by transition matrices of given order $L$ estimated on two DNA sequences of genes and intergenic regions.

\subsection{Randomly selected transition probabilities}


For each selection of $L$ and $K$, a symbol sequence of length $N$
is generated from a transition probability matrix of size $K^L
\times K$ with randomly selected components from the uniform
distribution $[0,1]$ under the restriction that the rows of the
matrix sum to one. In a pilot study we considered both the setting
of selecting a different transition probability matrix for each of
the 100 realizations and the setting of using the same transition
probability matrix for all realizations with different initial
conditions. The results were qualitatively the same and we chose
to proceed with the first setting.

\begin{figure}[htb]
\centerline{\hbox{\includegraphics[width=6cm]{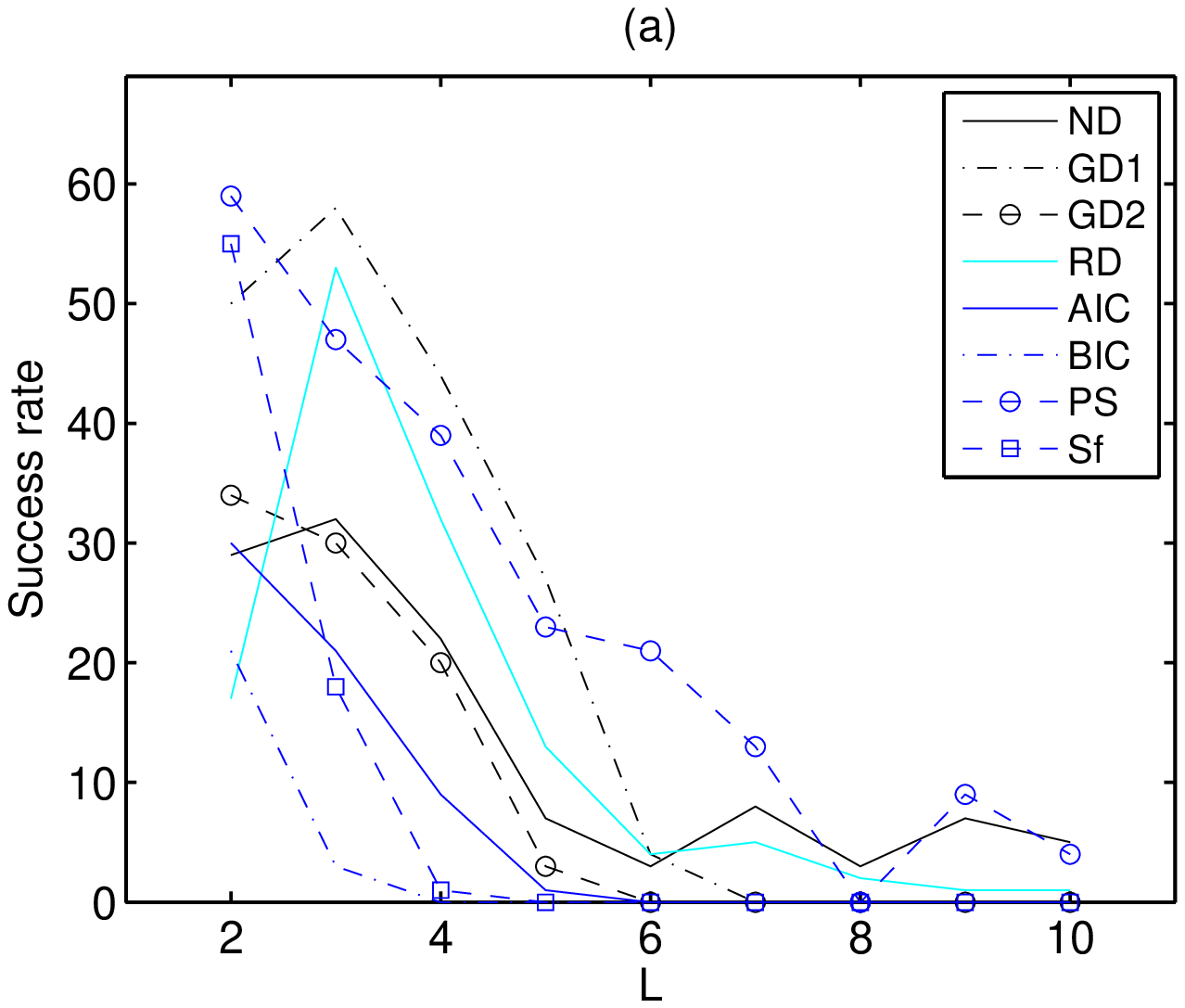}
\includegraphics[width=6cm]{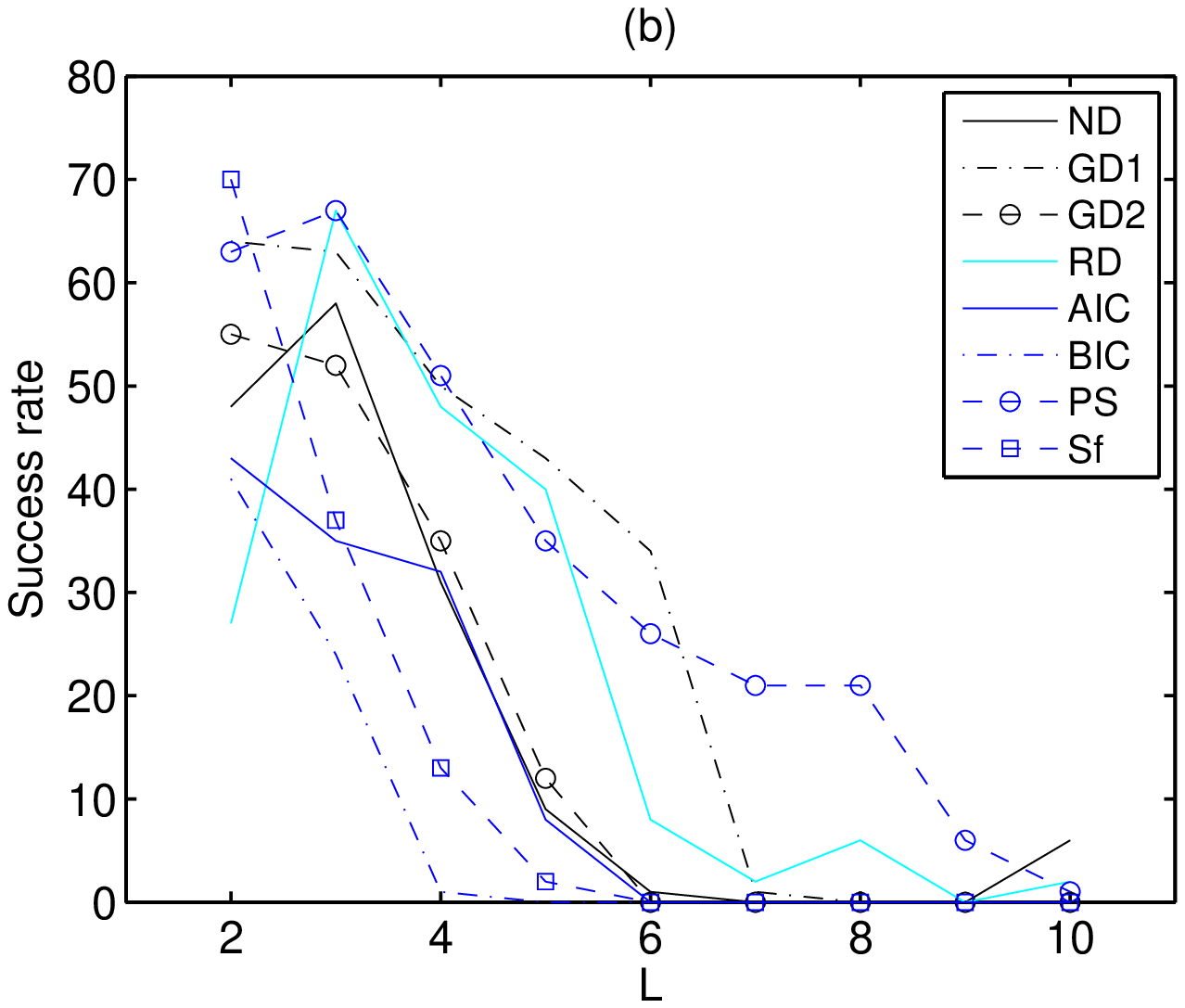}}}
\centerline{\includegraphics[width=6cm]{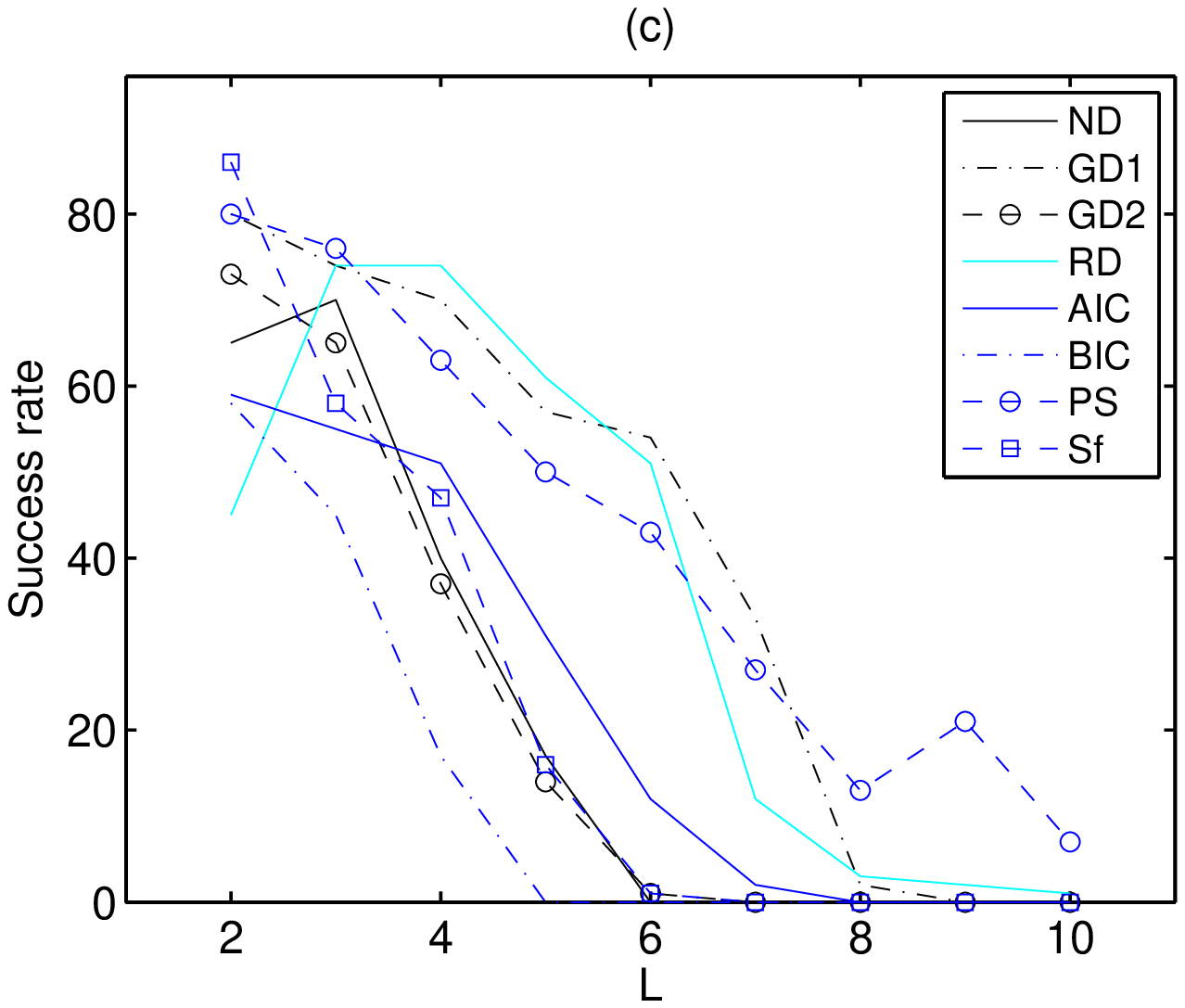}
\includegraphics[width=6cm]{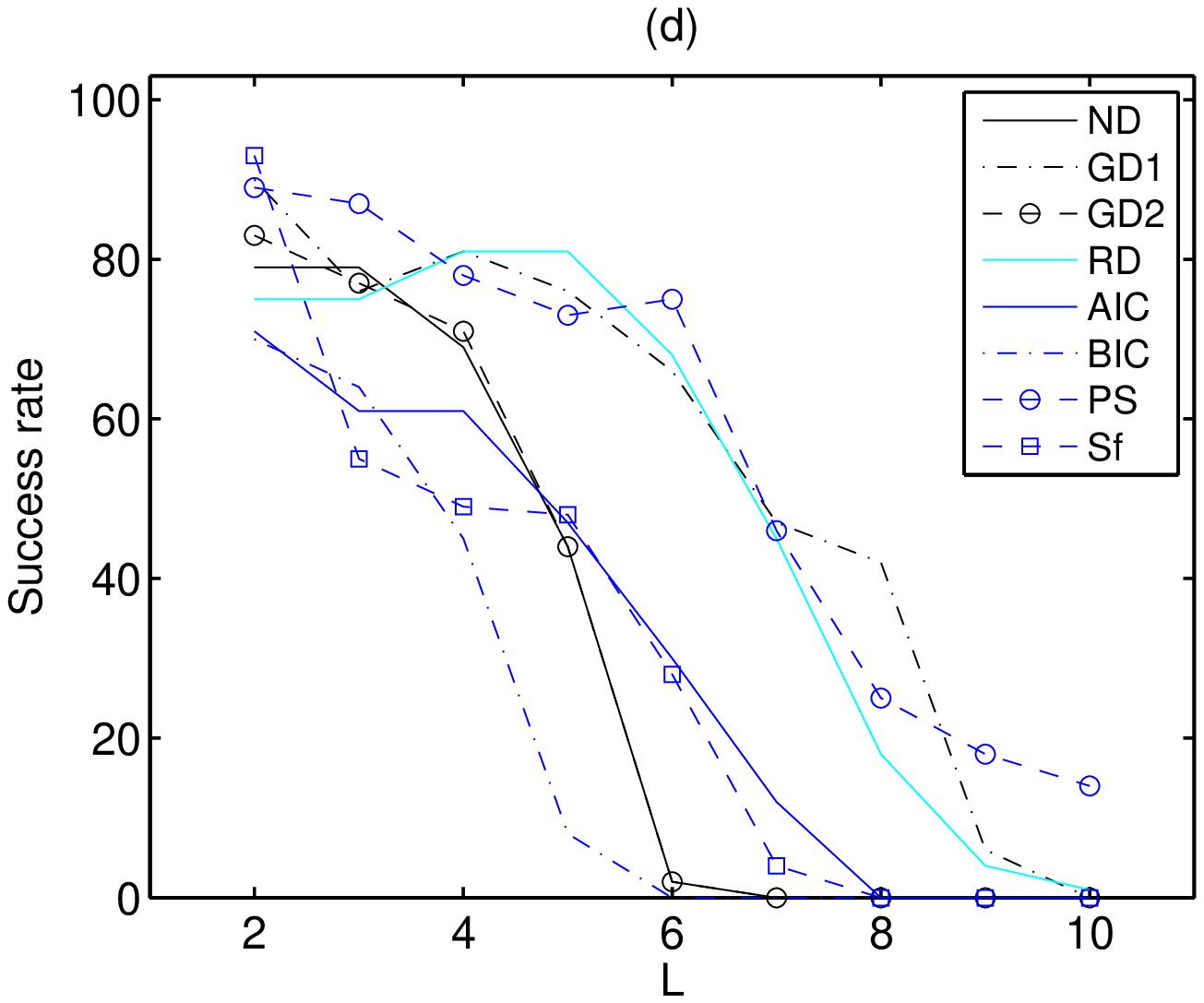}}
\centerline{\includegraphics[width=6cm]{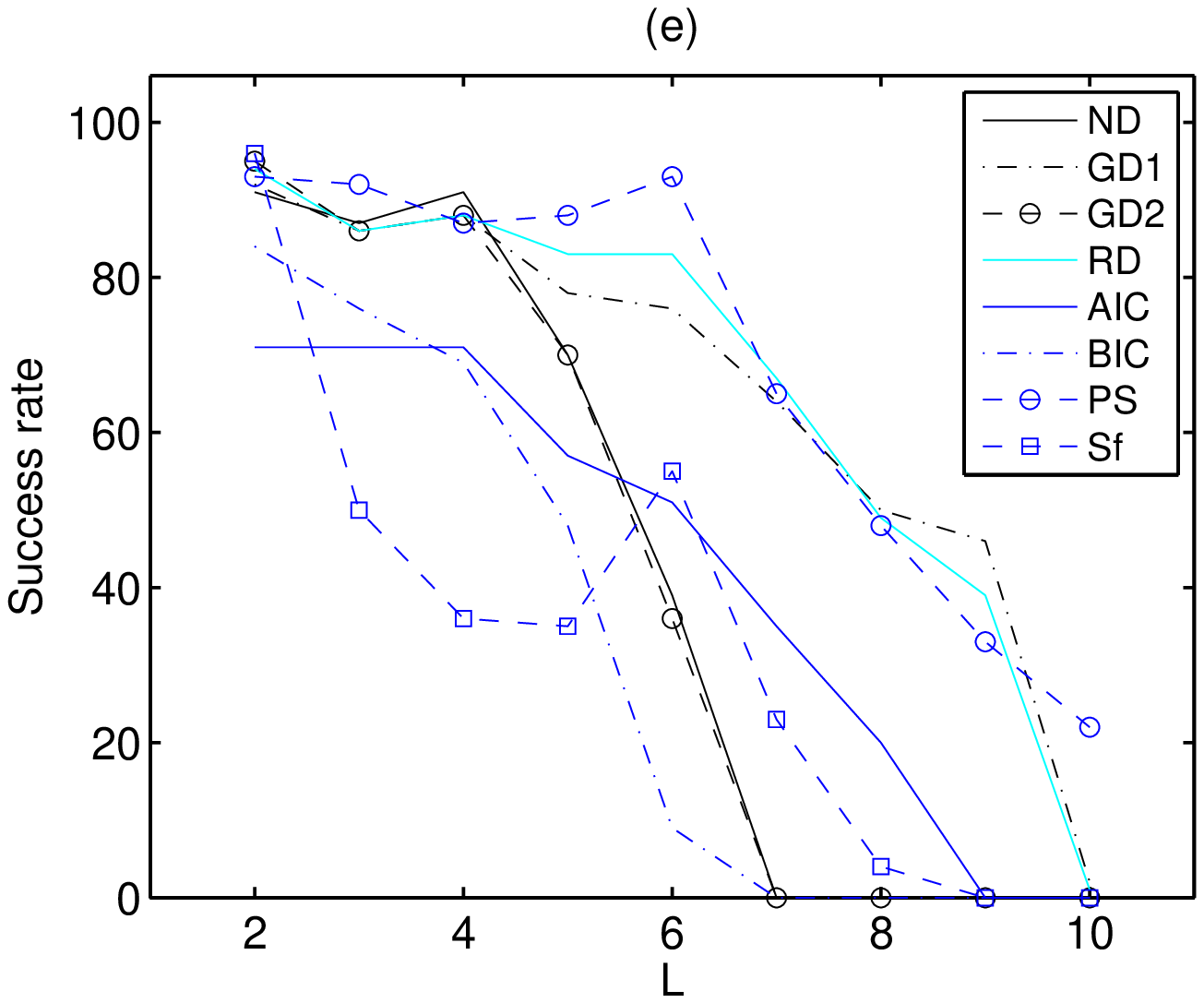}
\includegraphics[width=6cm]{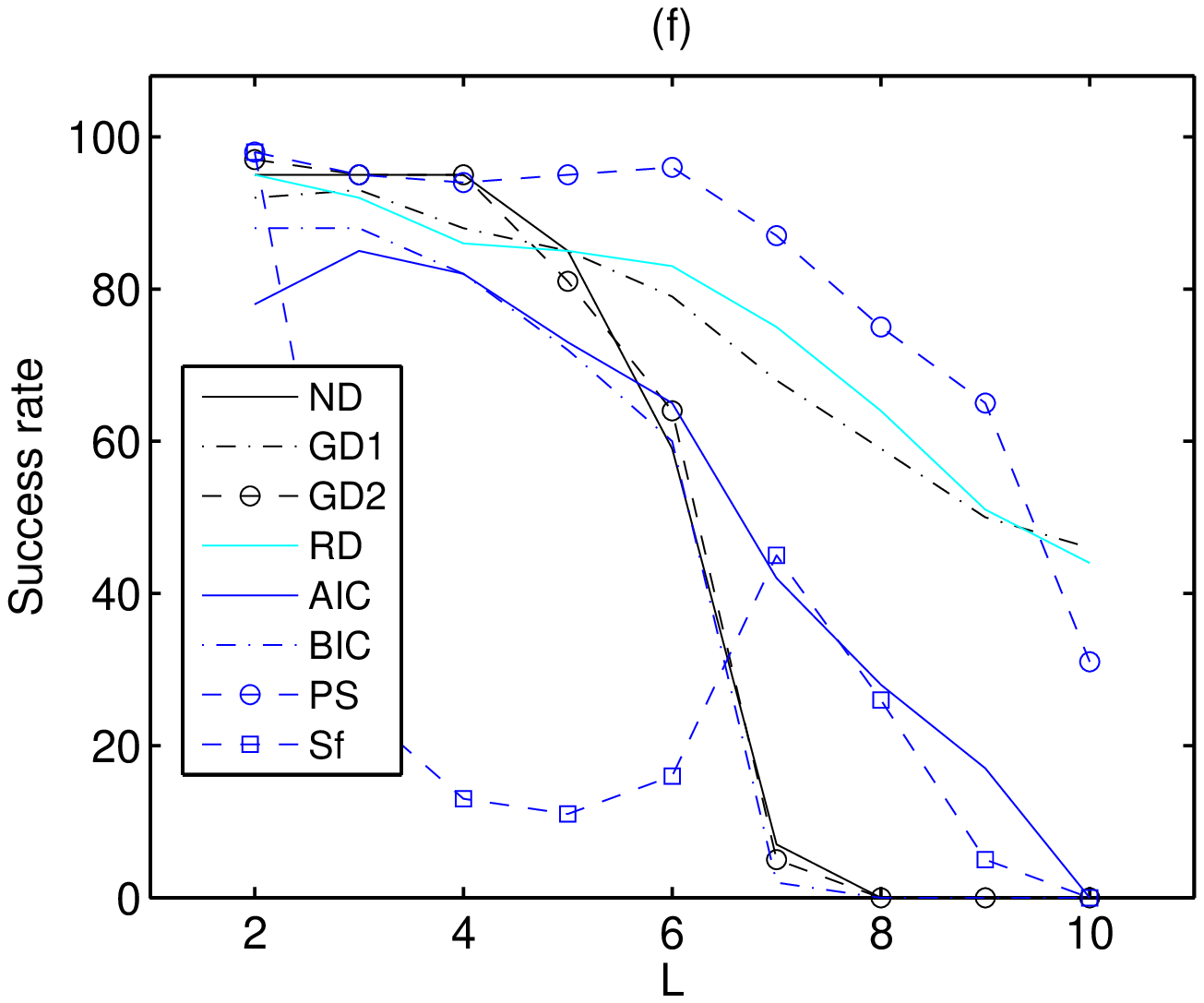}}
\caption{Number of cases out of $100$ realizations the true order $L$ is estimated by the criteria ND, GD1, GD2, RD, AIC, BIC, PS and Sf vs order
$L$, as shown in the legend. The symbol sequences have length (a) $N=100$, (b) $N=200$, (c) $N=400$, (d) $N=800$, (e) N=$1600$ (f) N=$3200$, and they
are generated by a Markov chain of $K=2$ symbols with a randomly selected transition probability matrix.} \label{fig:K2RandomMatrix}
\end{figure}

As expected, the simulations suggest that for all methods the
success rate in identifying the true order $L$ increases with $N$
and decreases with $L$ and $K$. As shown in
Figure~\ref{fig:K2RandomMatrix} for $K=2$, all criteria attain
about the same success rate in detecting the correct $L$ for
$L=2,3$ \textcolor{red}{. When $N \geq 1600$, the success rate
increases with $N$ being close to 100\% (see
Figure~\ref{fig:K2RandomMatrix}e and f). We can also notice that
the success rate} decreases with $L$ for any $N$ and for all
methods, but it decreases differently across the methods.
For each $N$ and as $L$ increases, the success rate of GD1, RD and PS decreases slower with $L$ than for the other criteria, with the success rate of
PS tending to stay positive even for $L=10$, e.g. see Figure~\ref{fig:K2RandomMatrix}c for $N=400$. It is worth noting that GD1 follows well with RD
for all $N$ and $L$ and at cases it even scores higher, e.g. for $N=200$ (Figure~\ref{fig:K2RandomMatrix}b) GD1 and RD have a success rate at about
40\% for $L=5$, while for $L=6$ the success rate decreases slightly for GD1 but dramatically for RD (the success rate of GD1 drops to zero for
$L=7$). In the same example, the success rate for PS decreases smoothly with $L$. For larger $N$ the three best criteria tend to align, and thus we
can safely conclude that these methods perform similarly and distinctly better than the other order estimation criteria. While all criteria improve
with $N$, Sf tends to score low even for small $L$.

\begin{figure}[htb]
\centerline{\hbox{\includegraphics[width=6cm]{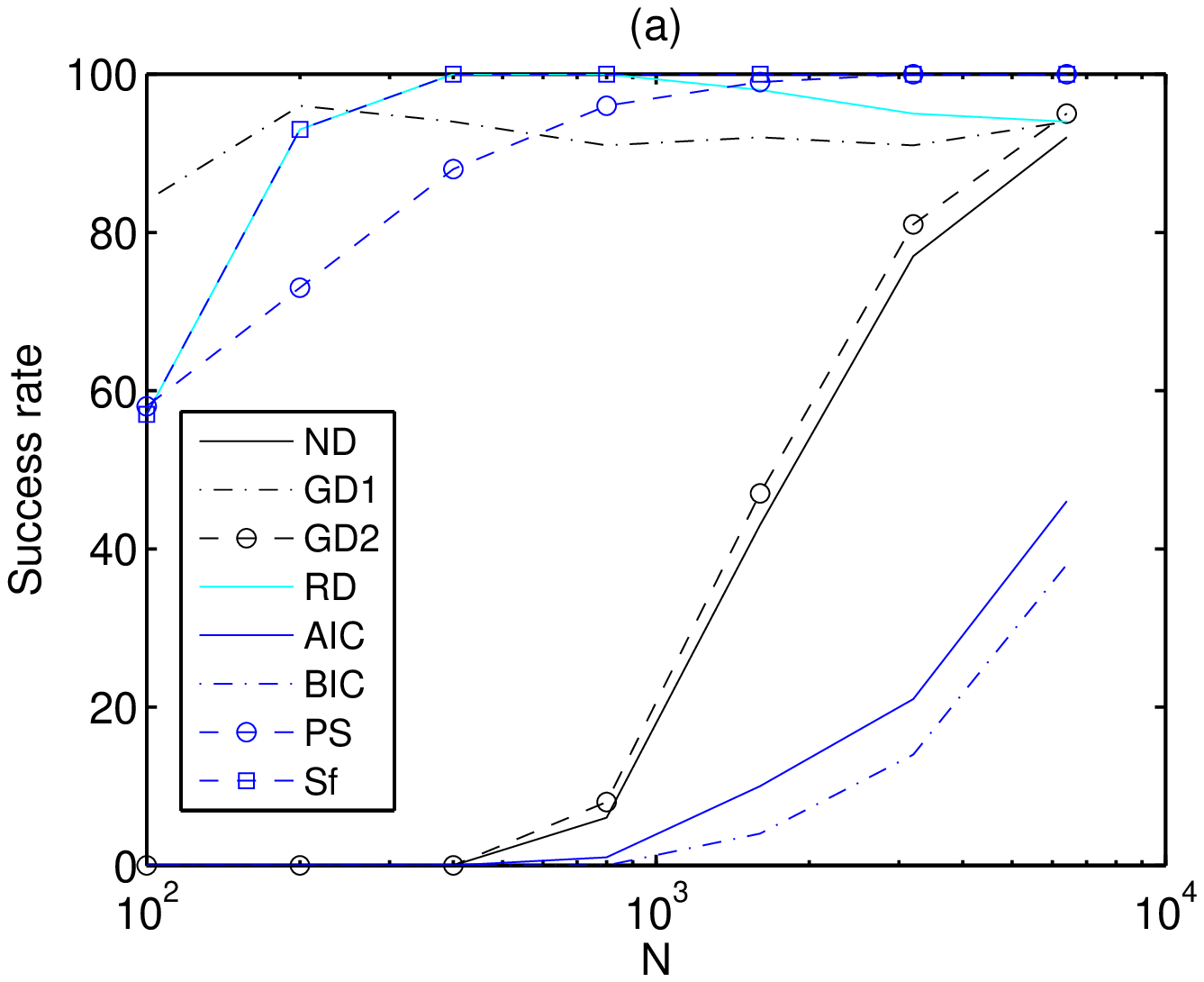}
\includegraphics[width=6cm]{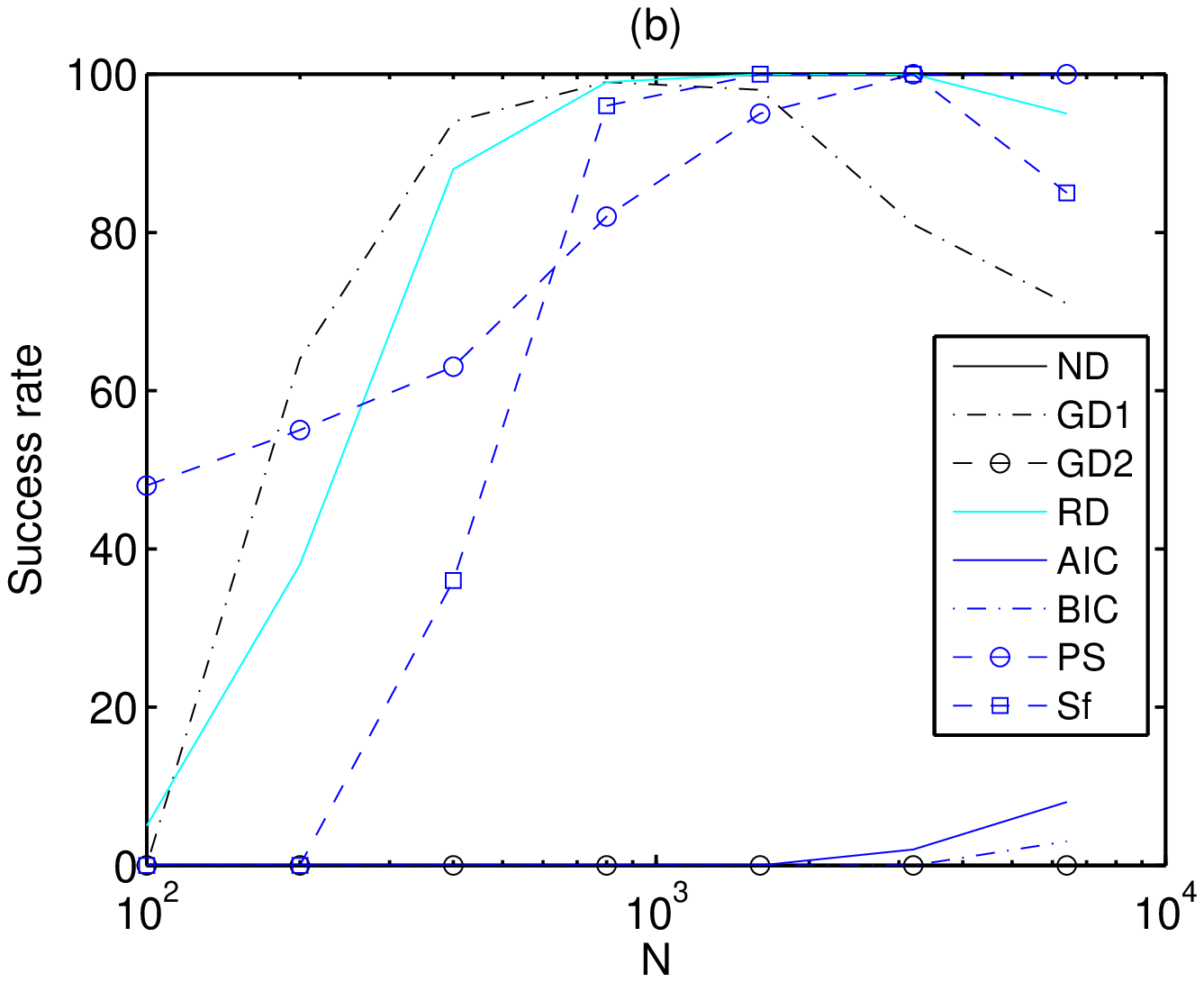}}}
\centerline{\includegraphics[width=6cm]{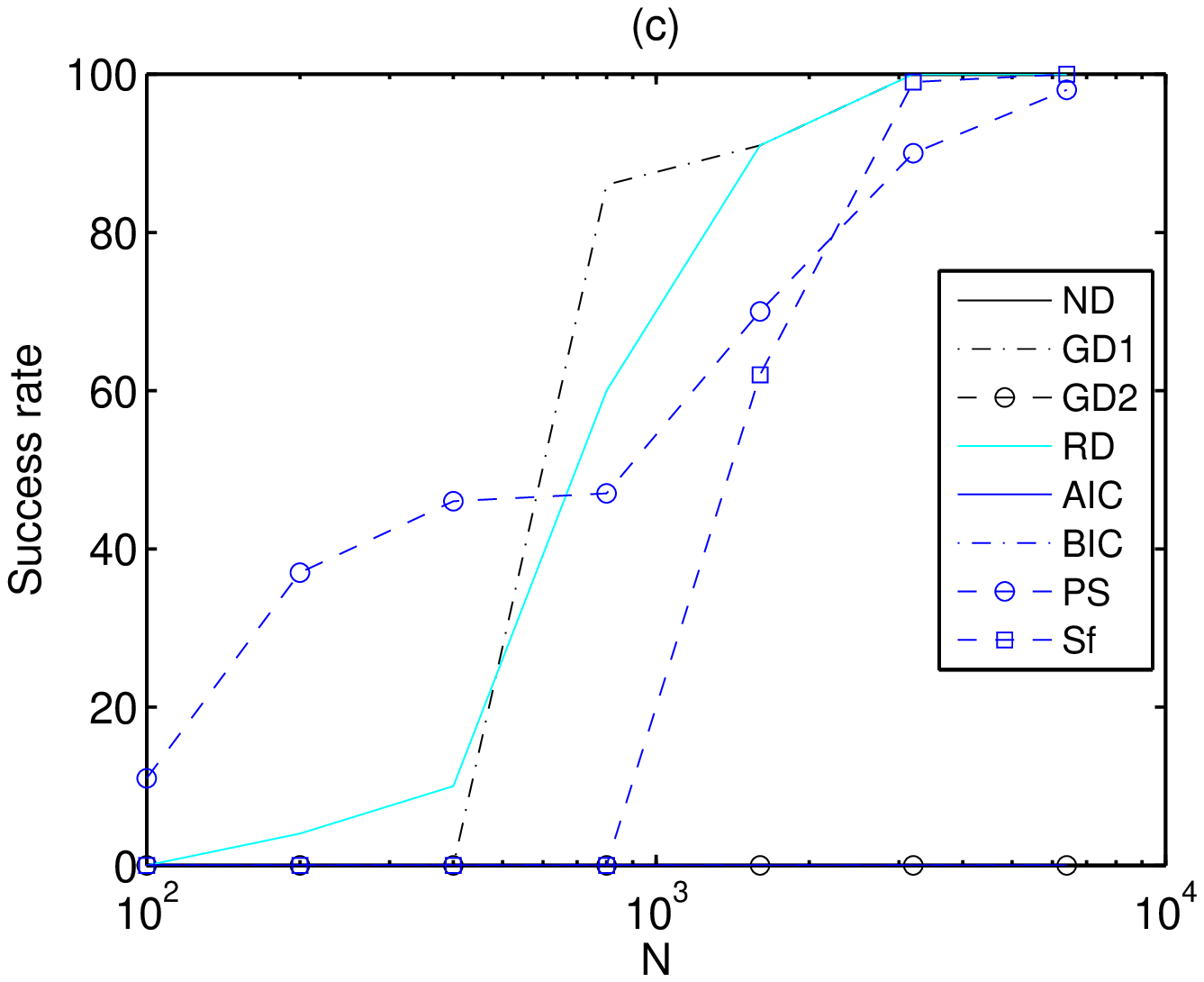}
\includegraphics[width=6cm]{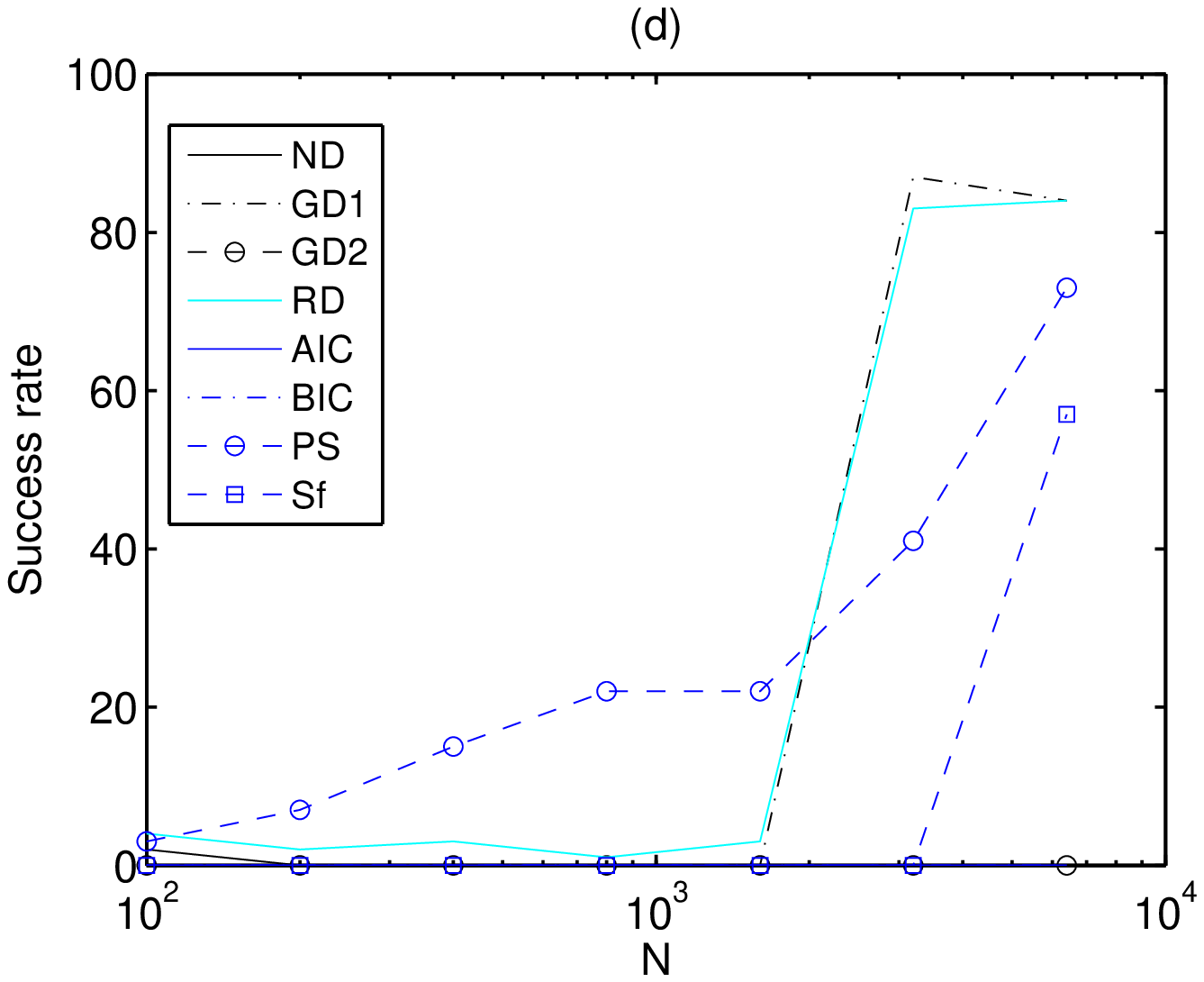}}
\caption{Number of cases out of $100$ realizations the true order $L$ is estimated by the criteria ND, GD1, GD2, RD, AIC, BIC, PS and Sf vs sequence
length $N$, as shown in the legend. The symbol sequences are generated by Markov chains of $K=4$ symbols with a randomly selected transition
probability matrix and order (a) $L=2$, (b) $L=3$, (c) $L=4$ and (d) $L=5$.} \label{fig:K4RandomMatrix}
\end{figure}

The estimation of $L$ is more data demanding as the number of
symbols increases, as shown in Figure~\ref{fig:K4RandomMatrix} for
$K=4$.
The success rate tends to increase with $N$, but for ND, GD2, AIC and BIC this can be seen only for small $L=2,3$ (Figure~\ref{fig:K4RandomMatrix}a
and b), while for larger $L=4,5$ (Figure~\ref{fig:K4RandomMatrix}c and d) even for the largest examined sequence length $N=6400$ the success rate is
zero. The three best criteria for $K=2$ perform also best for $K=4$ with Sf following close for small $L$ and scoring lower as $L$ increases. Here,
GD1 and RD have very similar performance, with GD1 scoring more often higher, and they both score highest in most cases, especially for large $L$ and
$N$.

\subsection{Transition probabilities estimated on DNA}

DNA consists basically of four nucleotides, the two purines,
adenine (A) and guanine (G), and the two pyrimidines, cytosine (C)
and thymine (T), so a DNA sequence can be considered as a symbol
sequence on the symbols A,C,G,T. In our analysis we use a large
segment of the Chromosome 1 of the plant A\emph{rabidopsis}
\emph{thaliana}\footnote{Data were obtained from the database:
http://www.ncbi.nlm.nih.gov}. We use two sequences, one sequence
derived by joining together the genes, which contain non-coding
regions, called introns, in between the coding regions, called
exons, and another sequence joining together the intergenic
regions which have solely non-coding character. The sequences used
here are segments of the long sequences used in
\cite{Kugiumtzis04c}.

\begin{figure}[htb]
\centerline{\hbox{\includegraphics[width=6cm]{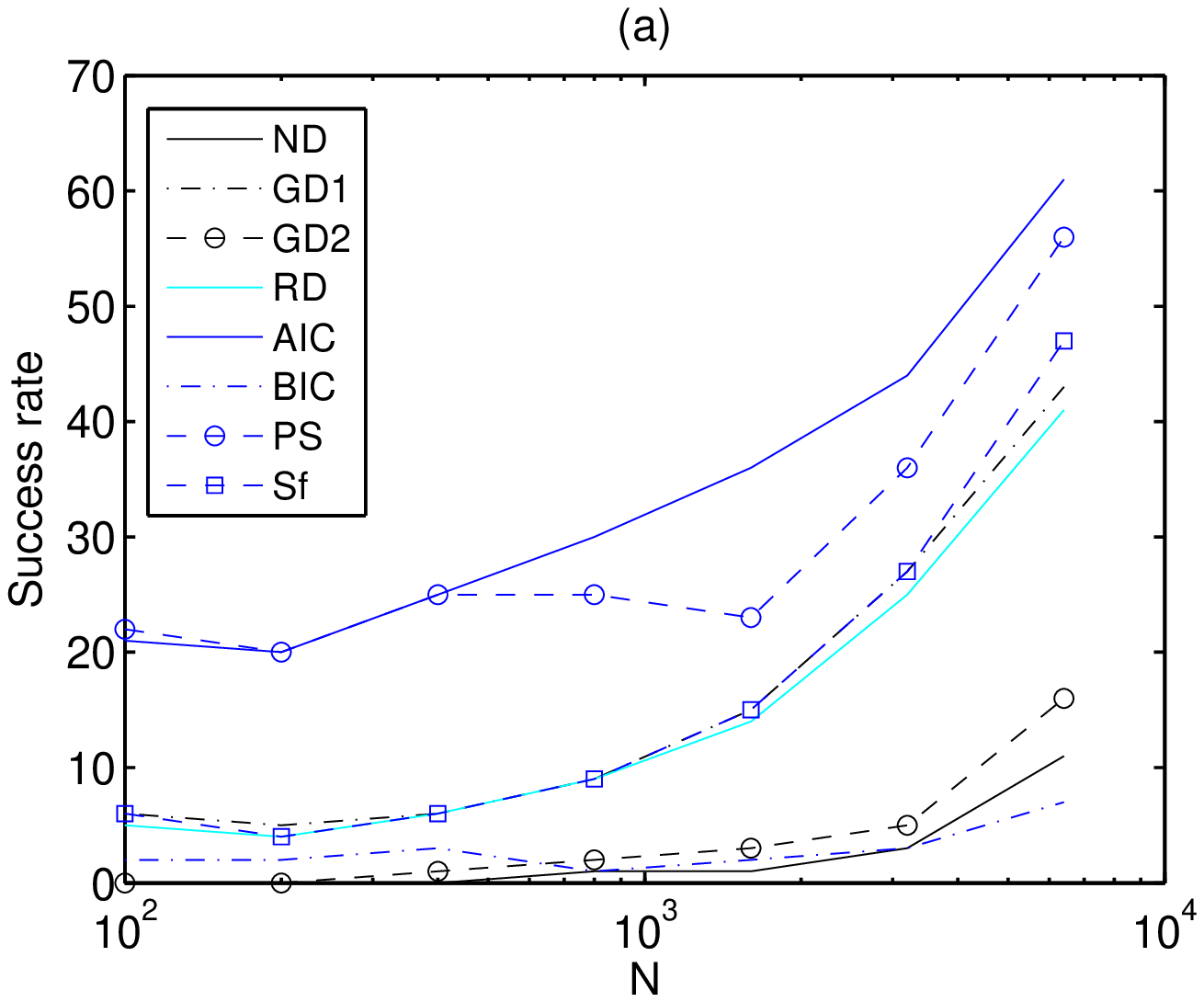}
\includegraphics[width=6cm]{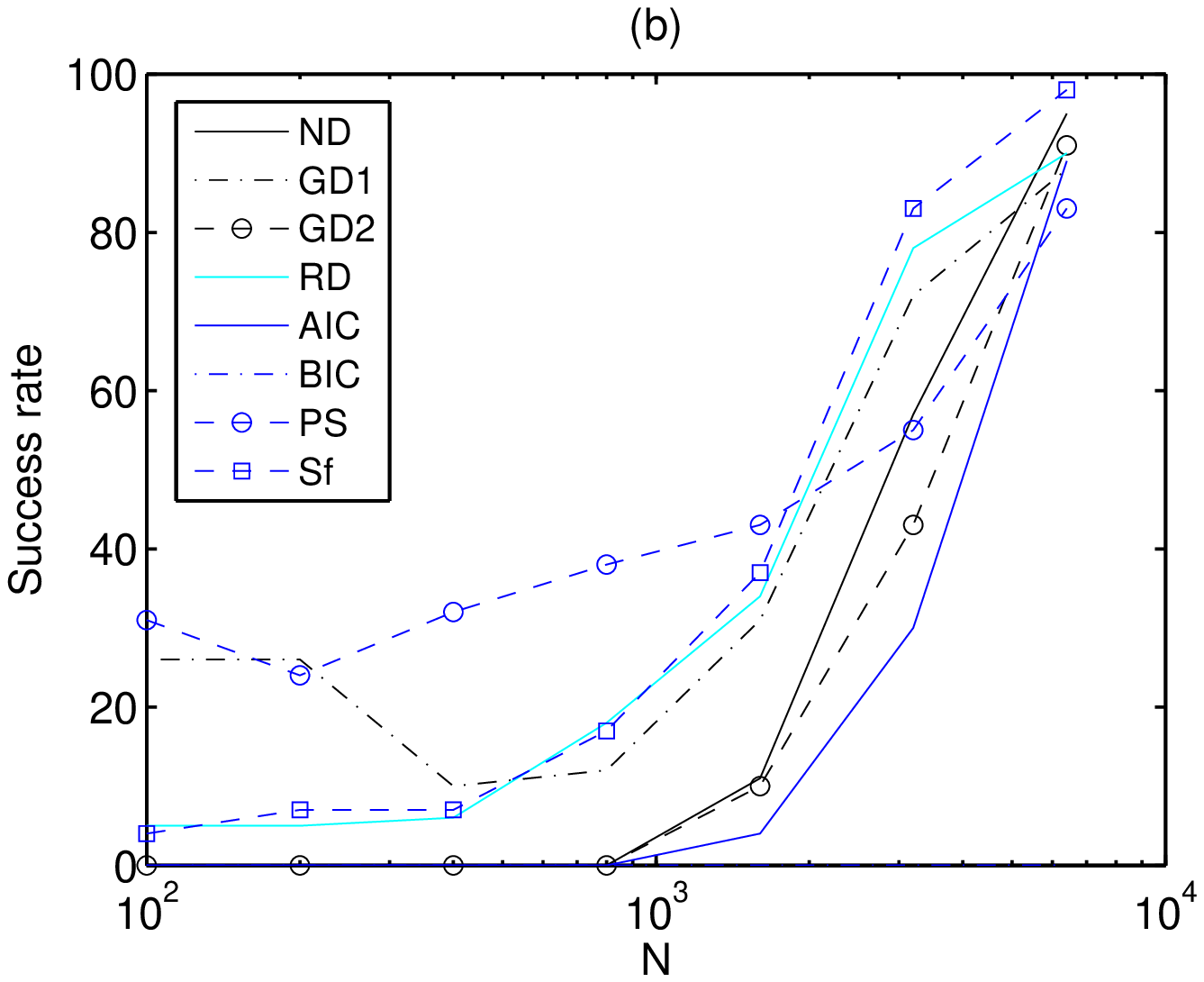}}}
\centerline{\includegraphics[width=6cm]{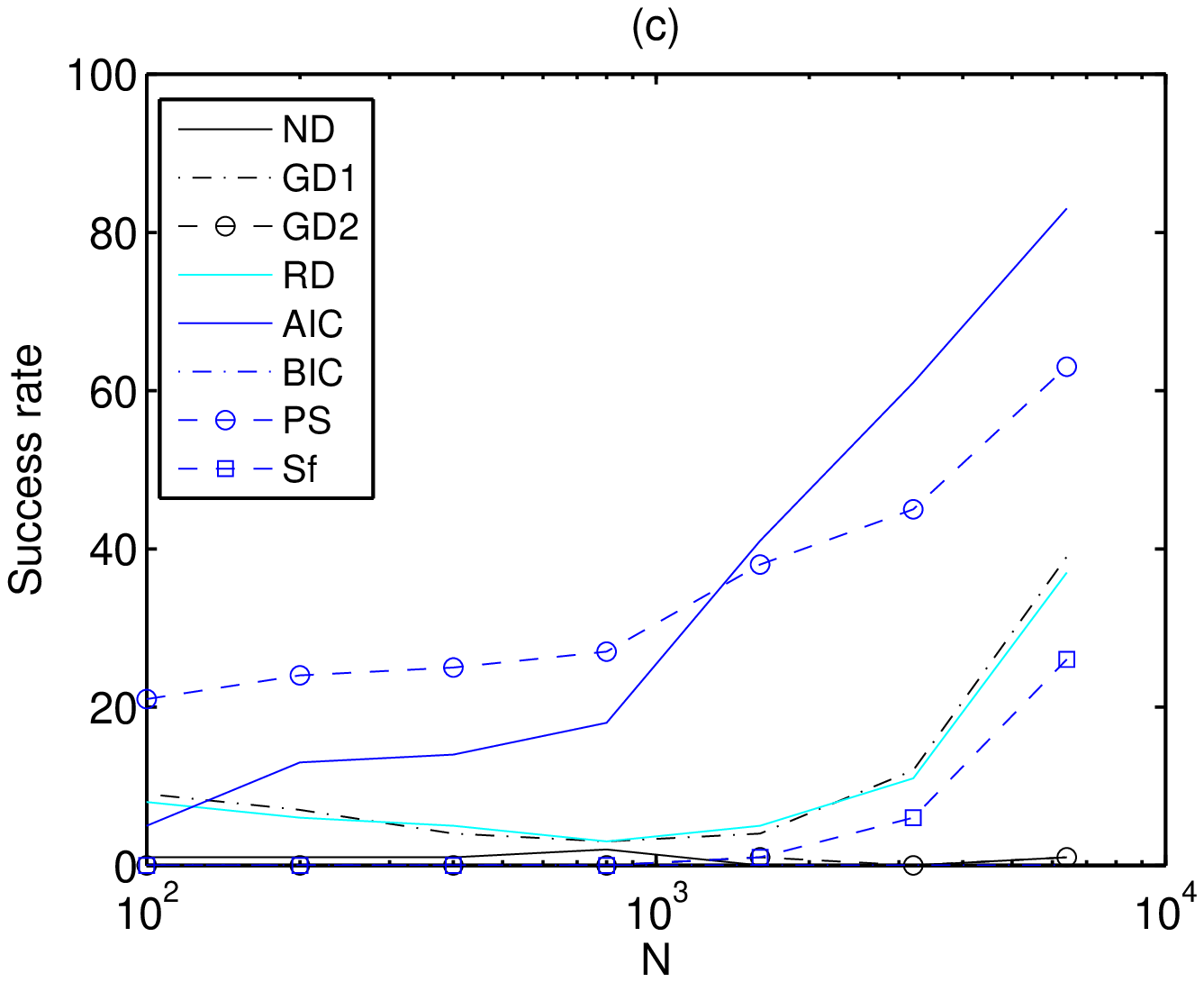}
\includegraphics[width=6cm]{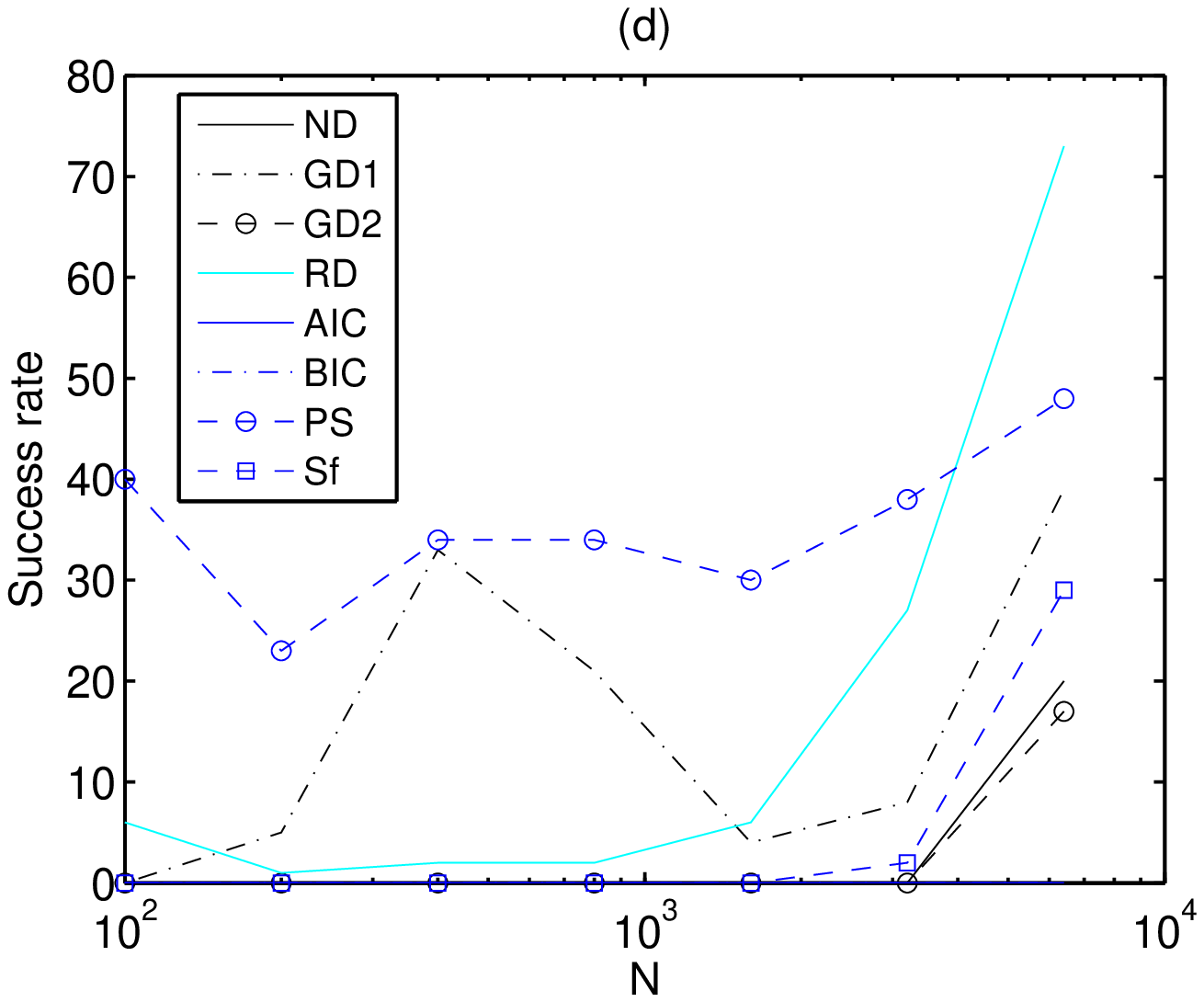}}
\centerline{\includegraphics[width=6cm]{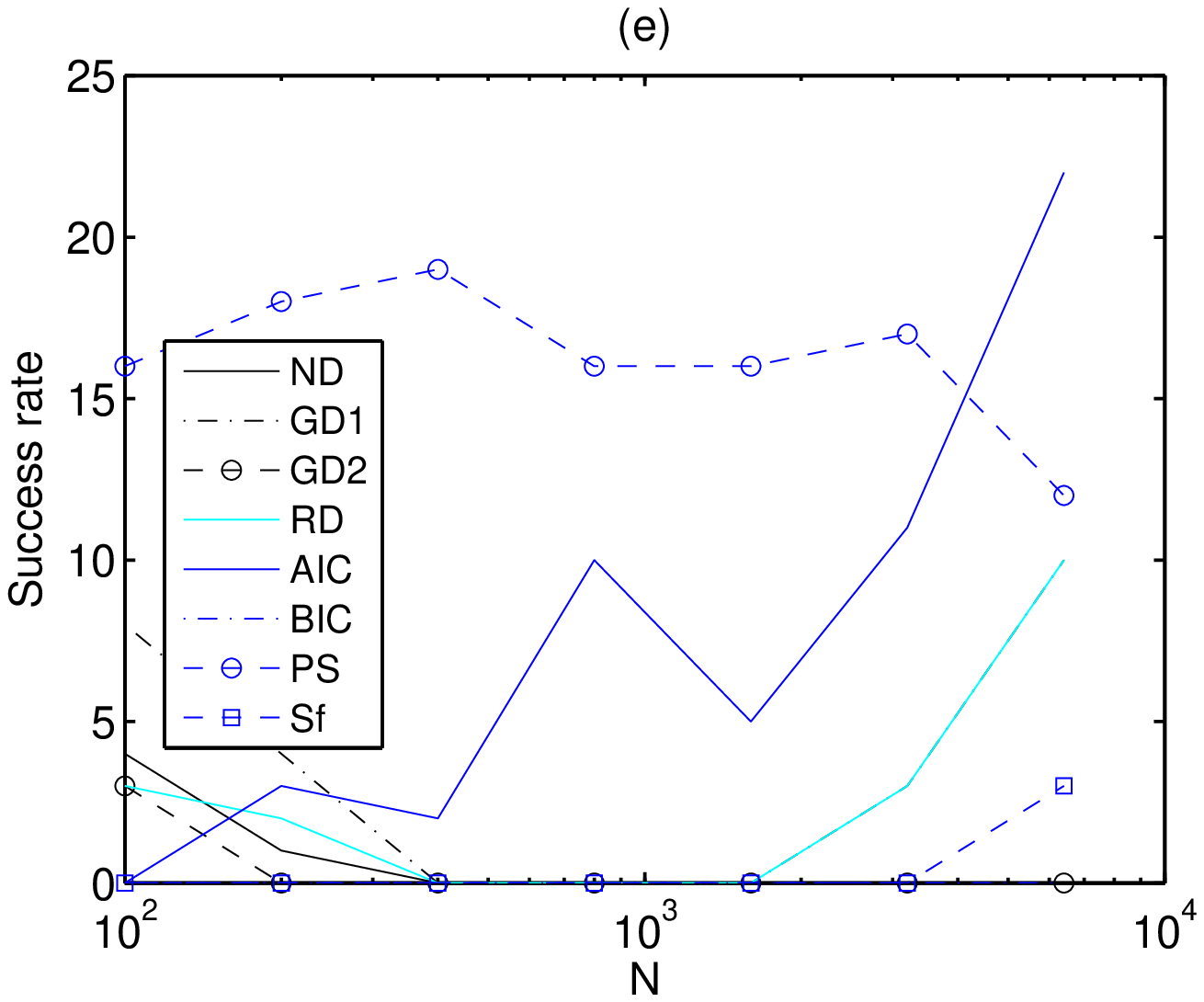}
\includegraphics[width=6cm]{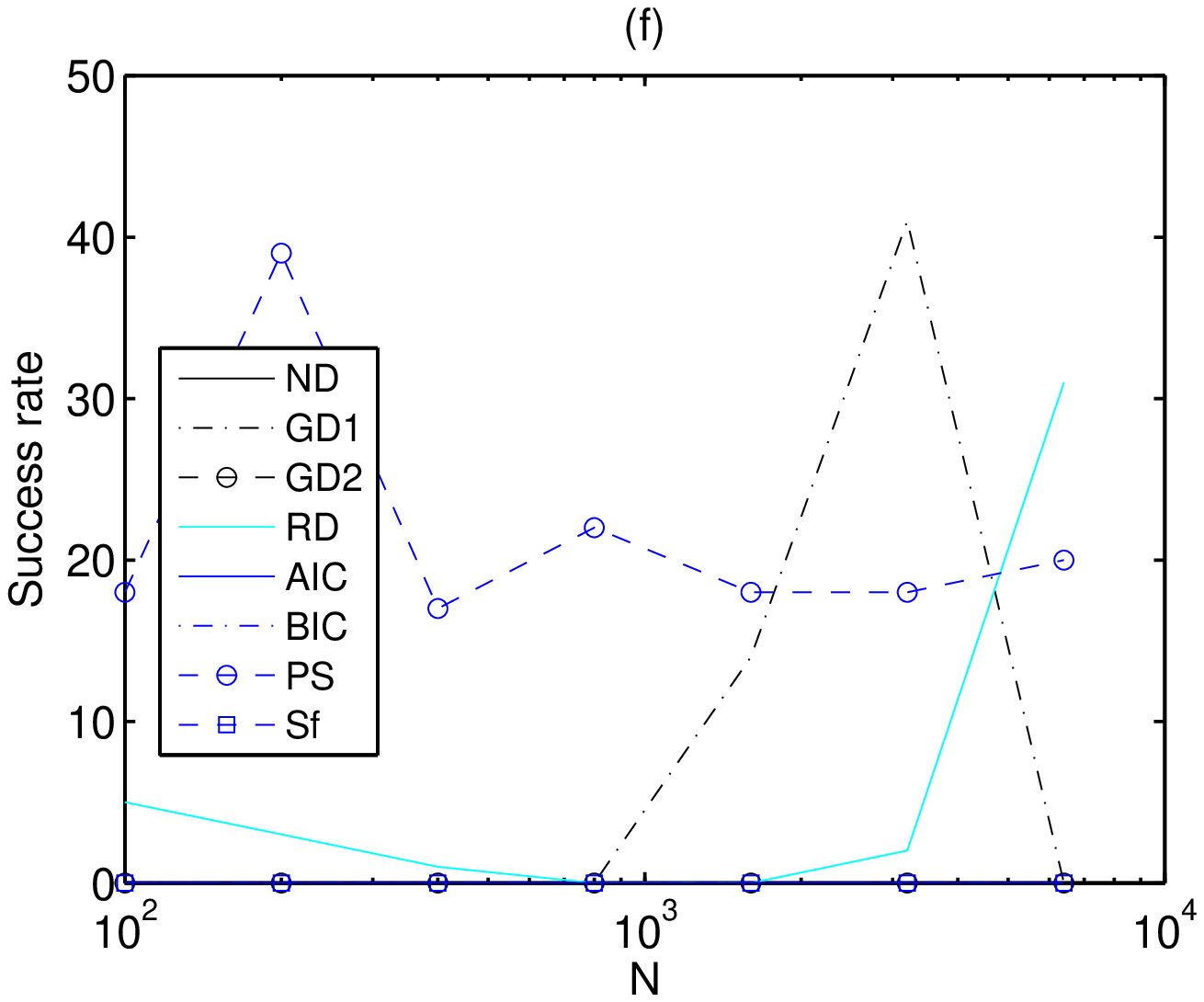}}
\centerline{\includegraphics[width=6cm]{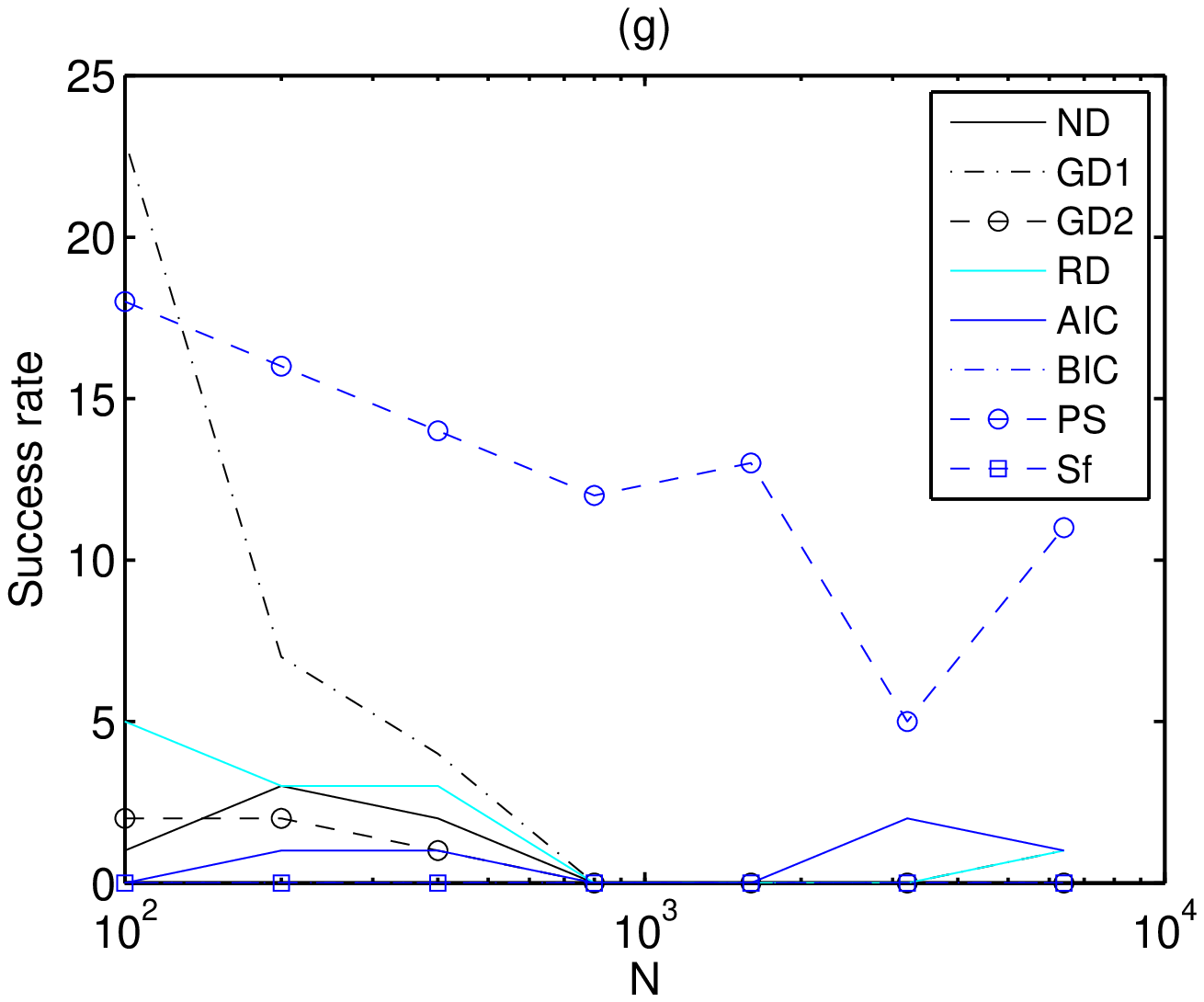}
\includegraphics[width=6cm]{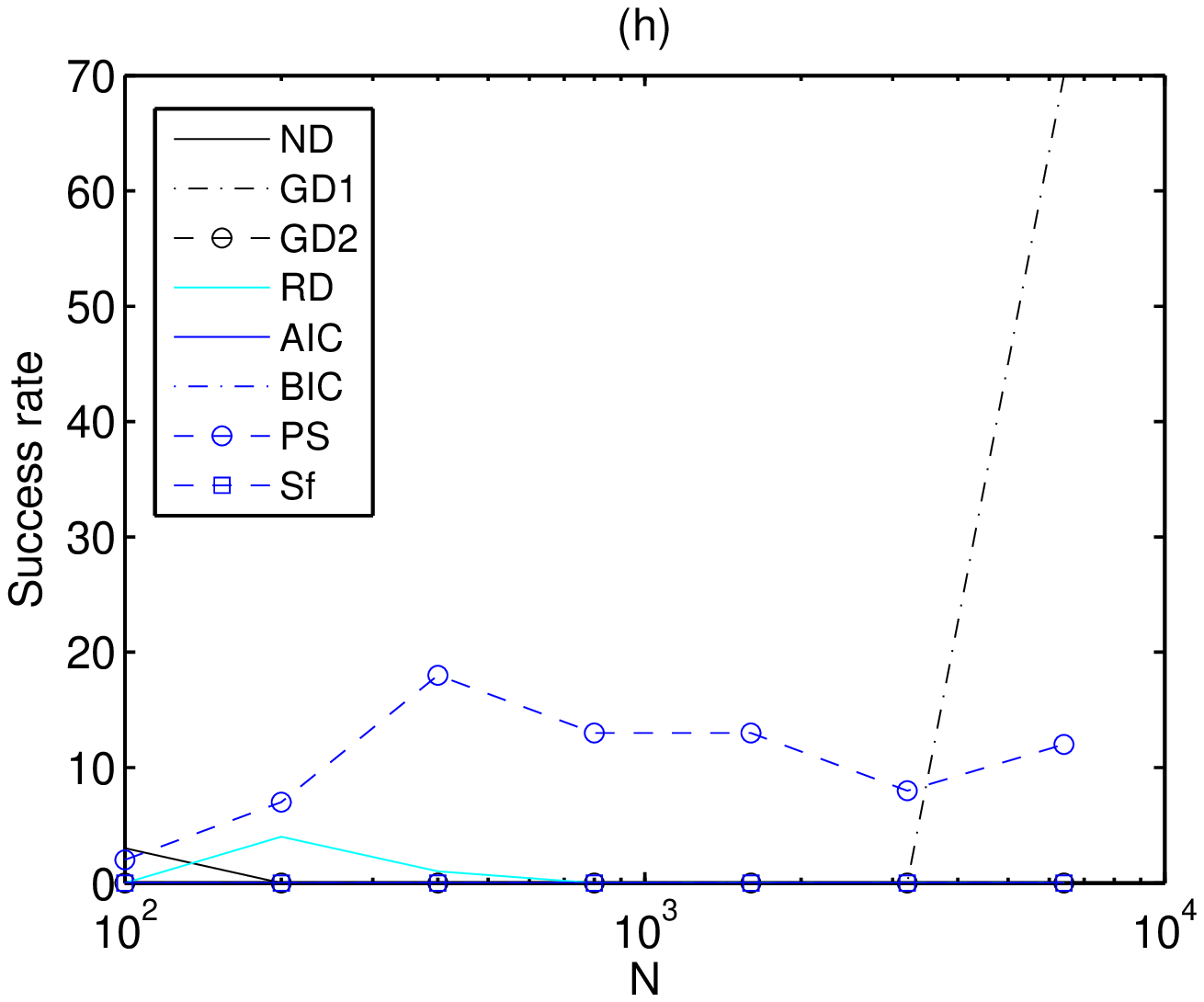}}
\caption{Number of cases out of $100$ realizations the true order $L$ is estimated by the criteria as shown in the legend, vs sequence length $N$.
The symbol sequences are generated by Markov chains of transition probability matrices estimated on a DNA sequence of genes. The panels are for
purines and pyrimidines ($K=2$) and $L=2,3,4,5$ in (a), (c), (e), (g), respectively, and for the four nucleotides ($K=4$) and $L=2,3,4,5$ in panels
(b), (d), (f), (h), respectively.} \label{fig:KLPMapogarC}
\end{figure}

\begin{figure}[htb]
\centerline{\hbox{\includegraphics[width=6cm]{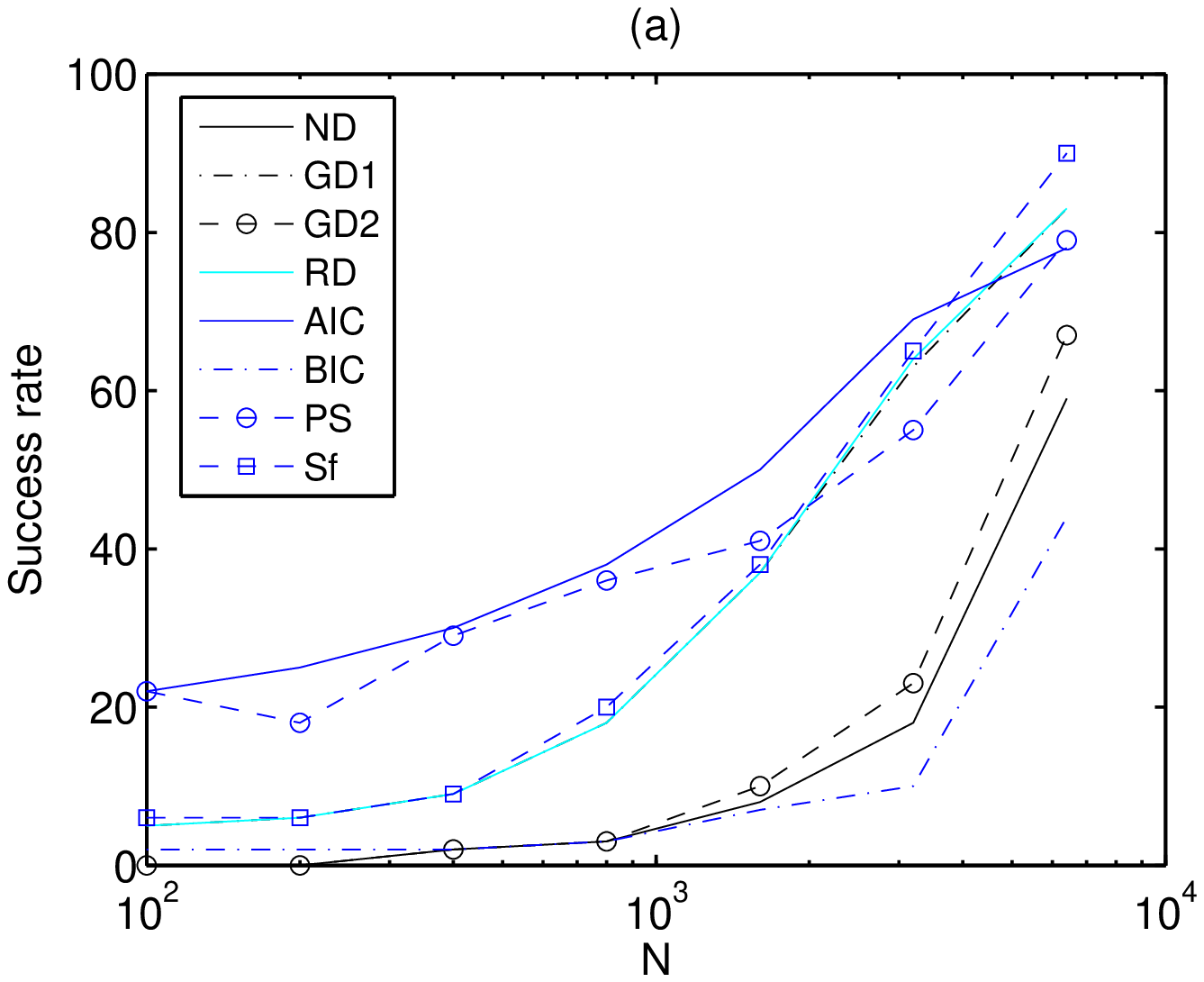}
\includegraphics[width=6cm]{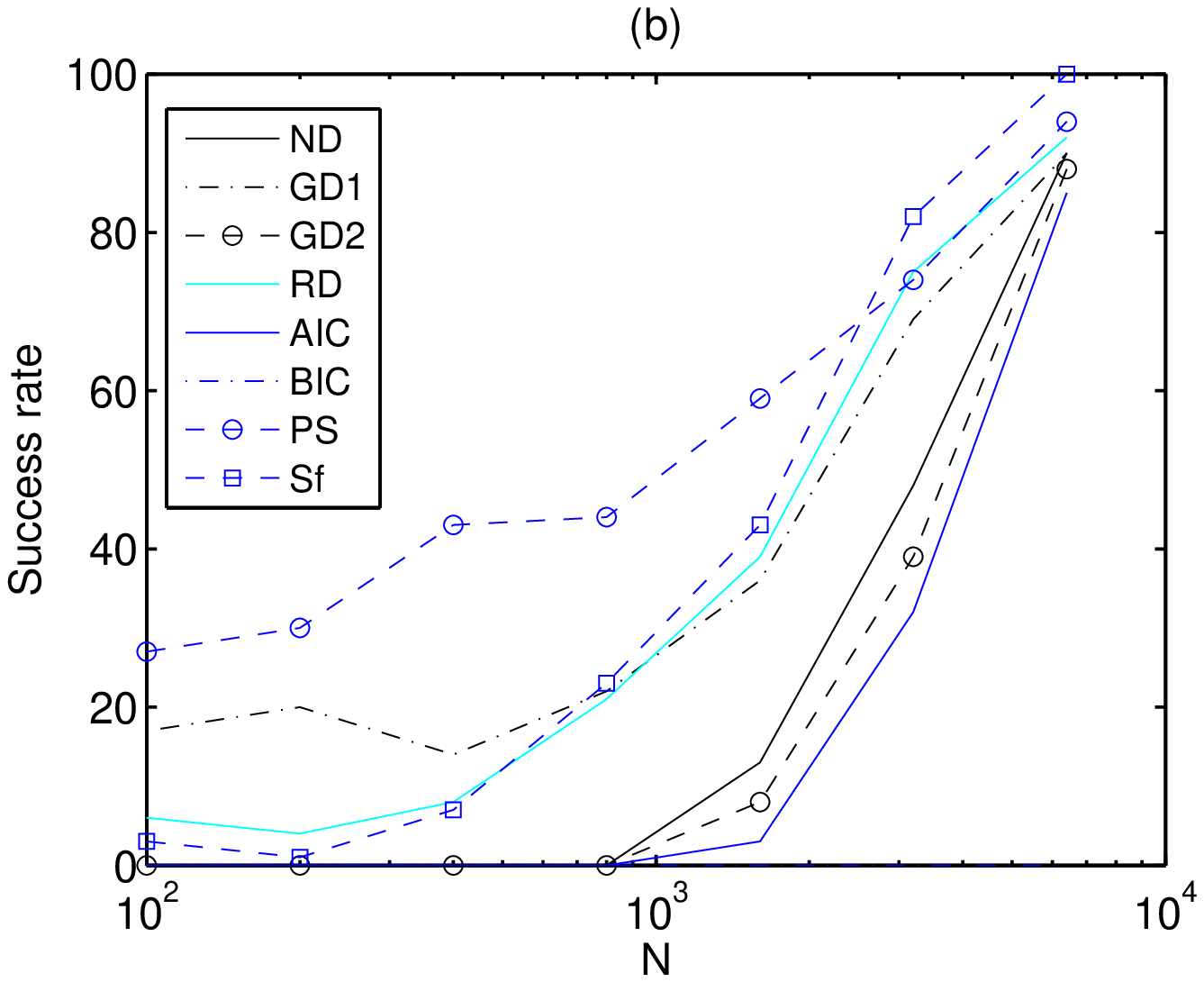}}}
\centerline{\includegraphics[width=6cm]{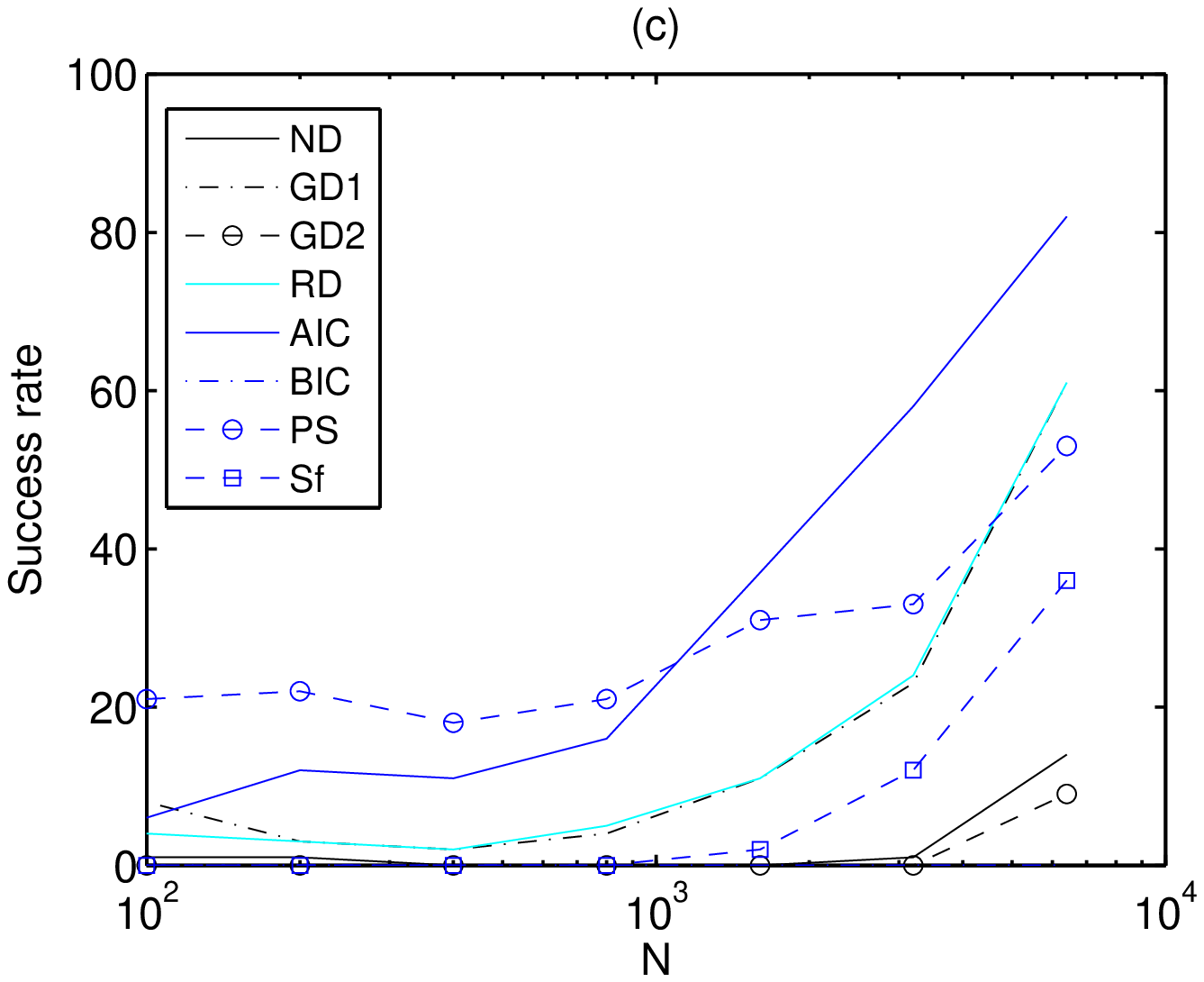}
\includegraphics[width=6cm]{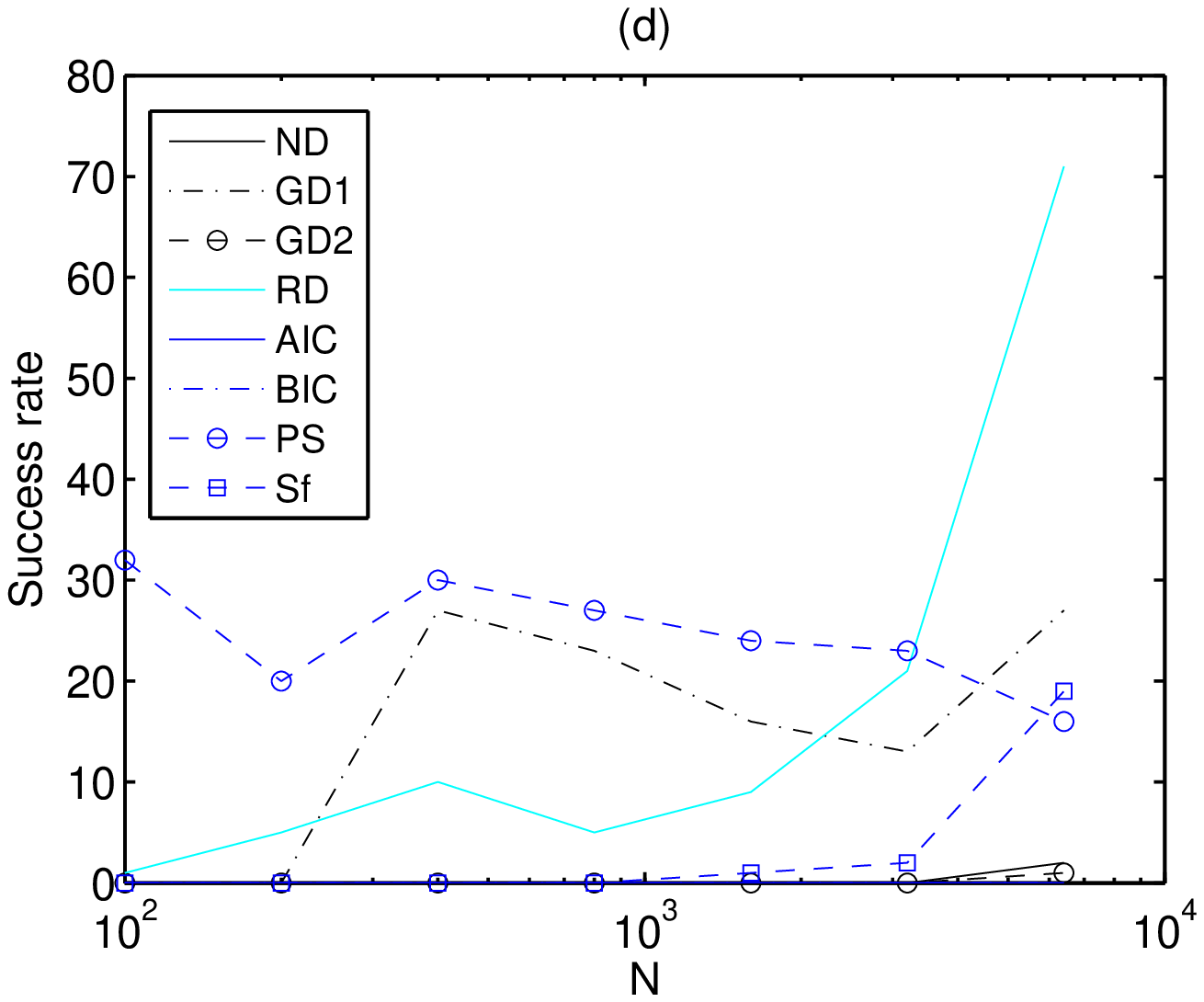}}
\centerline{\includegraphics[width=6cm]{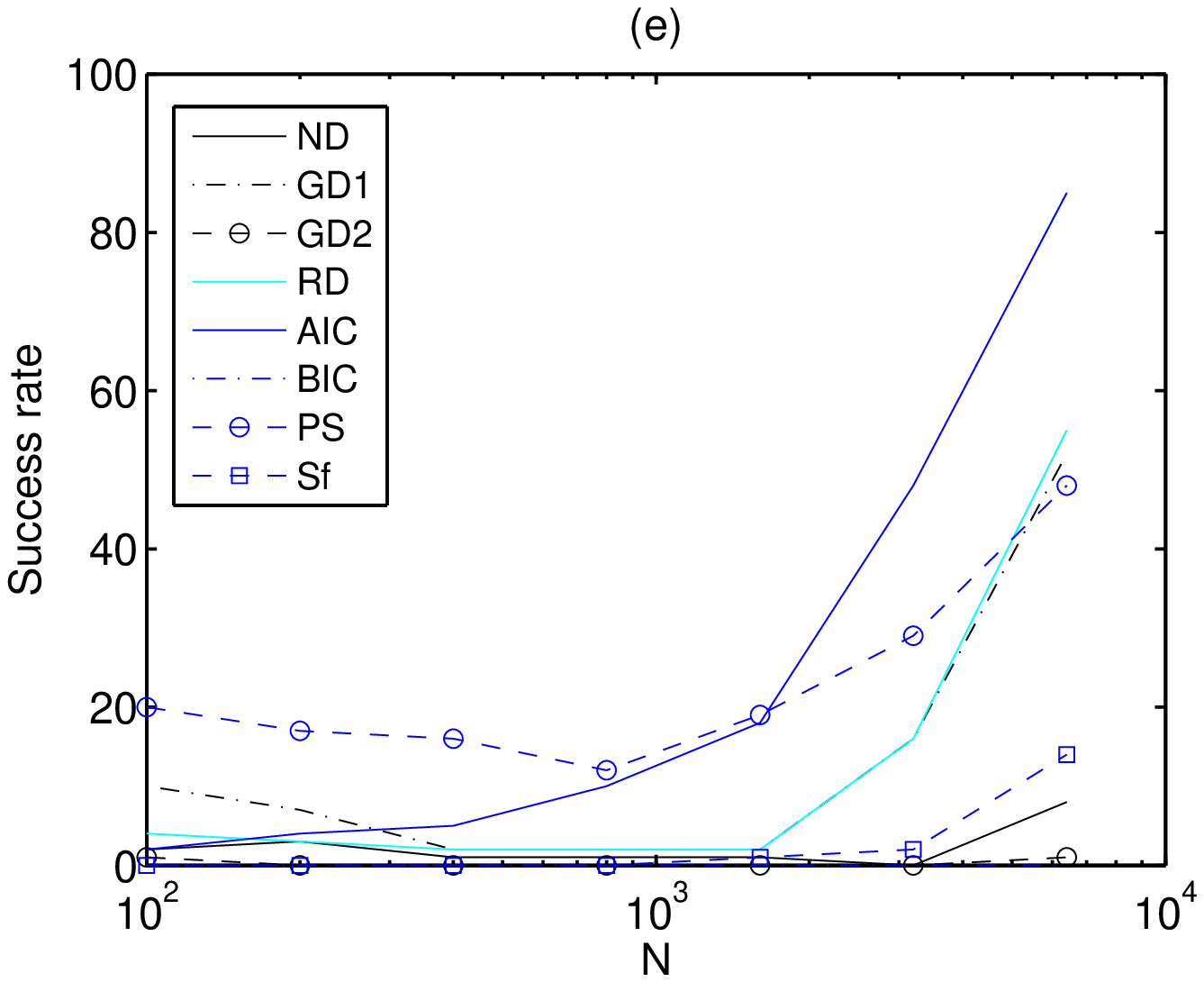}
\includegraphics[width=6cm]{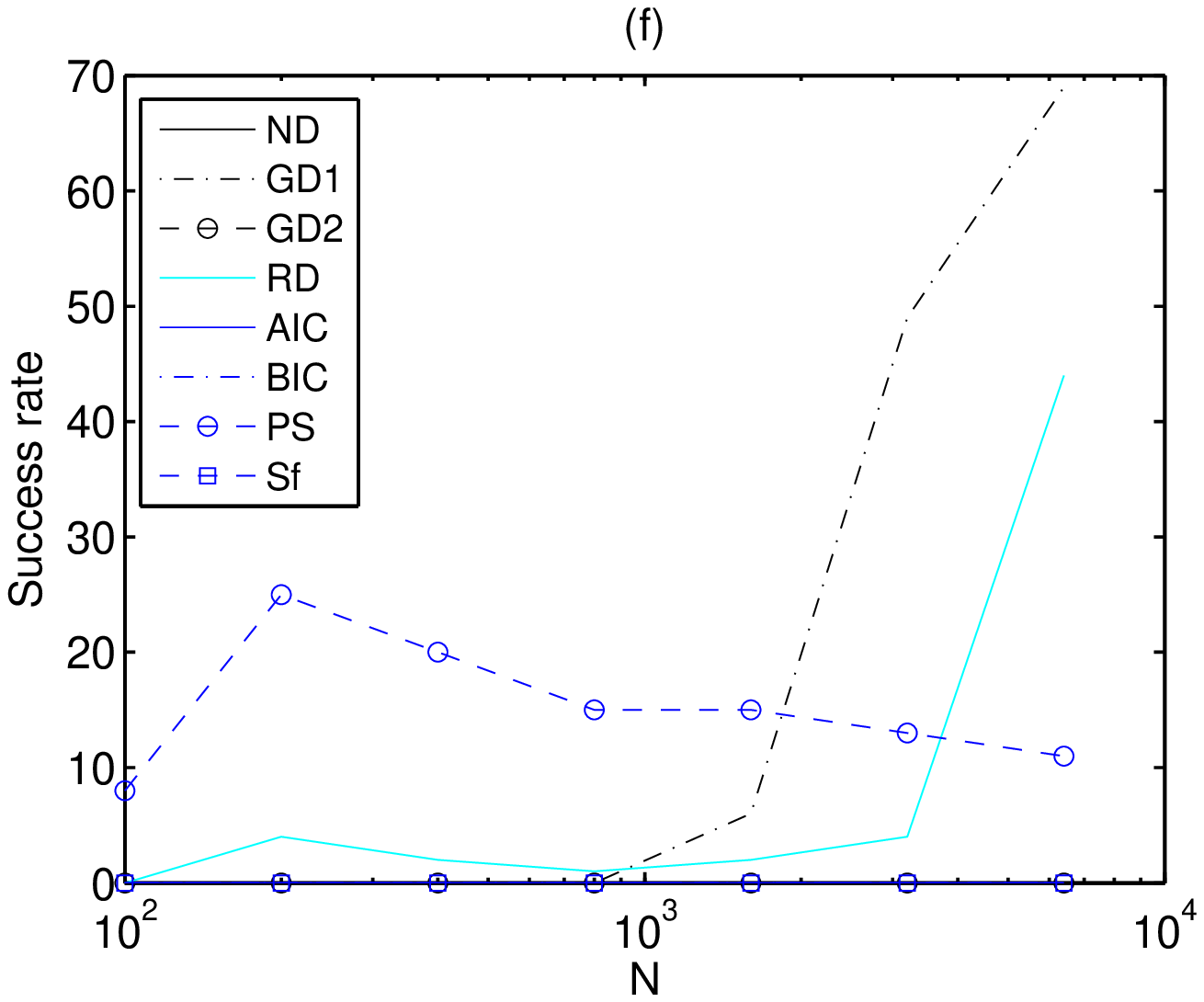}}
\centerline{\includegraphics[width=6cm]{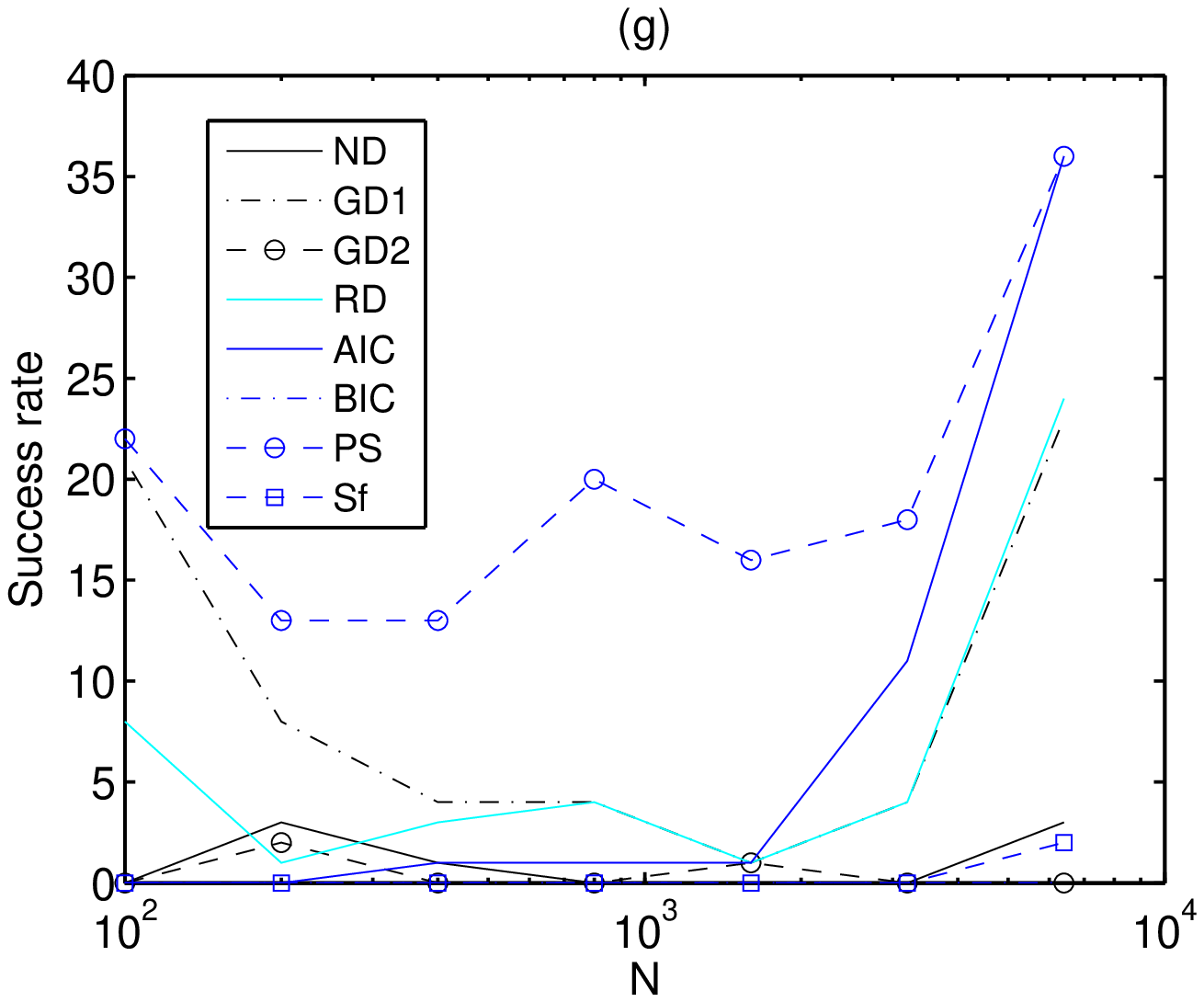}
\includegraphics[width=6cm]{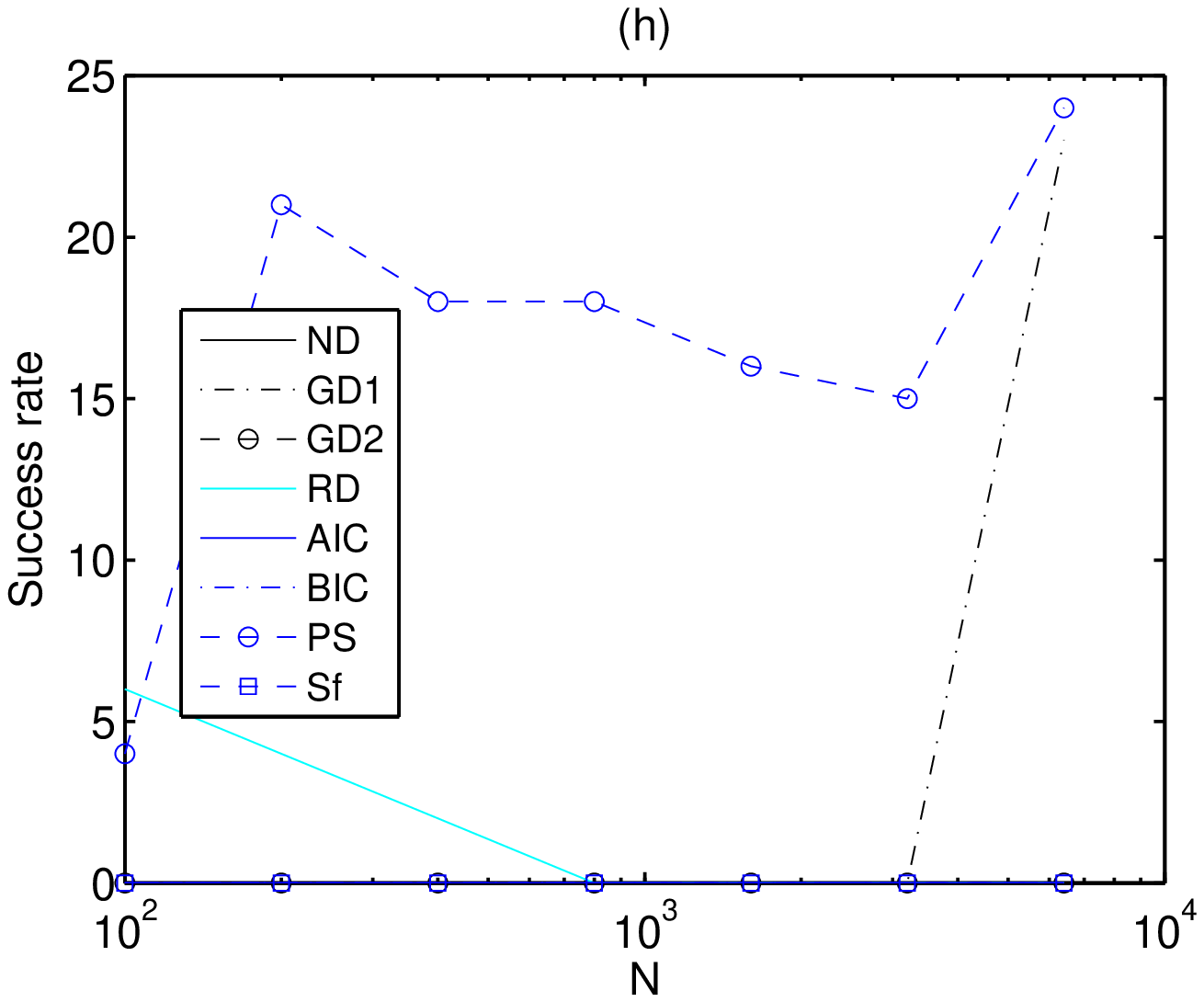}}
\caption{(a) Same as for Figure~\ref{fig:KLPMapogarC}, but for the DNA sequence of intergenic regions.} \label{fig:KLPMapogarNC}
\end{figure}

In this simulation setup we form the Markov chains from transition
probabilities matrices of given order $L$ estimated on the two DNA
sequences of genes and intergenic regions, and we generate 100
symbol sequences from each of these Markov chains for different
initial conditions. The purpose here is to consider Markov chains of
distinct structure of the probability transition matrix for each order
$L$ that relate to a real world Markov chain.
The results for the success rate of correct
estimation of the true order $L$ with all the criteria and for
$K=2$ (purine and pyrimidine) and $K=4$ (all four nucleotides),
where we set $L=2,3,4,5$ and $N$ varying from 100 to 6400, are
shown in Figure~\ref{fig:KLPMapogarC} for the genes and in
Figure~\ref{fig:KLPMapogarNC} for the intergenic regions.
For both genes and intergenic regions, all the criteria fail when the order gets large ($L>3$) and only PS maintains a positive success rate but at
the same low level of 10\% - 20\% and rather independently of $N$. For smaller orders ($L=2,3$), all criteria tend to improve with $N$ but at low
levels of success rate differing across the criteria (for $L=2$ see Figure~\ref{fig:KLPMapogarC}a and b for genes and $K=2$ and $K=4$, respectively,
and the same in Figure~\ref{fig:KLPMapogarNC}a and b for intergenic regions). These results suggest that the task of estimating the true $L$ of a
Markov chain with the structure of transition probabilities as in DNA sequences is more difficult than when the transition probabilities are selected
at random. Concerning the CMI-based tests, again ND and GD2 fail to estimate the true $L$ for both genes and intergenic regions, while GD1 follows
tightly with RD, both being suboptimal but scoring consistently well compared to all other criteria. For example for genes and $L=2$, when $K=2$
(Figure~\ref{fig:KLPMapogarC}a) GD1 and RD score lower than PS and AIC for all $N$ (and higher than all others), but when $K=4$
(Figure~\ref{fig:KLPMapogarC}b) GD1 and RD score higher than AIC for all $N$ and PS at large $N$. AIC scores highest of all criteria for $K=2$ but it
has zero success rate when $K=4$, and only for $L=2$ the success rate increases above zero with large $N$ (Figure~\ref{fig:KLPMapogarC}b), indicating
that the data requirement for AIC with the increase of $K$ is disproportionately high compared to the other criteria. On the other hand, PS estimates
correctly the order $L$ at the same low rate regardless of $N$ for $L>3$, being however higher than for other criteria. This somehow peculiar
performance of PS is explained by the fact that for $L>3$ PS estimates at random the order $L$, so that it hits the true order at a percentage of
cases dependent on the range of the tested $m$ values, whereas the other criteria underestimate the order. GD1 and RD have thus the most consistent
behavior, increasing the probability (success rate) to identify the true order with $N$ at a level depending on $L$ and $K$.

Comparing the results of the criteria for the two types of DNA sequences, they match pretty well for the corresponding $K$, $L$ and $N$. Though the
relative differences of the criteria are the same, the level of success rate tends to be higher for the intergenic regions, specifically for $K=2$,
indicating that the Markov chain of the same order $L$ obtained on the basis of intergenic regions is less complex, i.e. the order is better
detectable than for the genes. For example,
 for $K=2$ and $L=3$, it can be seen in
 Figure~\ref{fig:KLPMapogarC}c that GD1 and RD reach
a success rate of 40\% at the largest tested $N=6400$ for genes, while for the intergenic regions the corresponding success rate is at 60\%
(Figure~\ref{fig:KLPMapogarNC}c). The overall results show that when the transition probabilities are estimated on DNA sequences of genes and
intergenic regions, all the criteria fail for larger orders, having somehow higher success rates for intergenic regions.

\section{Application on DNA sequences}
\label{sec:DNA}
Much of the statistical analysis of DNA sequences is focused on the estimation of properties of coding and non-coding regions as well as on the
discrimination of these regions. There has been evidence that there is a different structure in coding and non-coding sequences and that the
non-coding sequences tend to have long range correlation, whereas the correlation in coding sequences exhibits exponential decay
\citep{Peng92,Buldyrev98,Almirantis99}. Here we use intergenic and gene sequences. The latter is a mixture of coding regions (exons) and non-coding
regions (introns), and therefore we expect to have also long range correlation due to the non-coding regions in it, but it should be less than the
correlation in the intergenic regions consisting only of non-coding parts. Thus both DNA sequences cannot be considered as Markov chains, at least
not of a moderate order, and the estimation of the order $L$ should increase with the available data size.

\begin{figure}[htb]
\centerline{\hbox{\includegraphics[width=7cm]{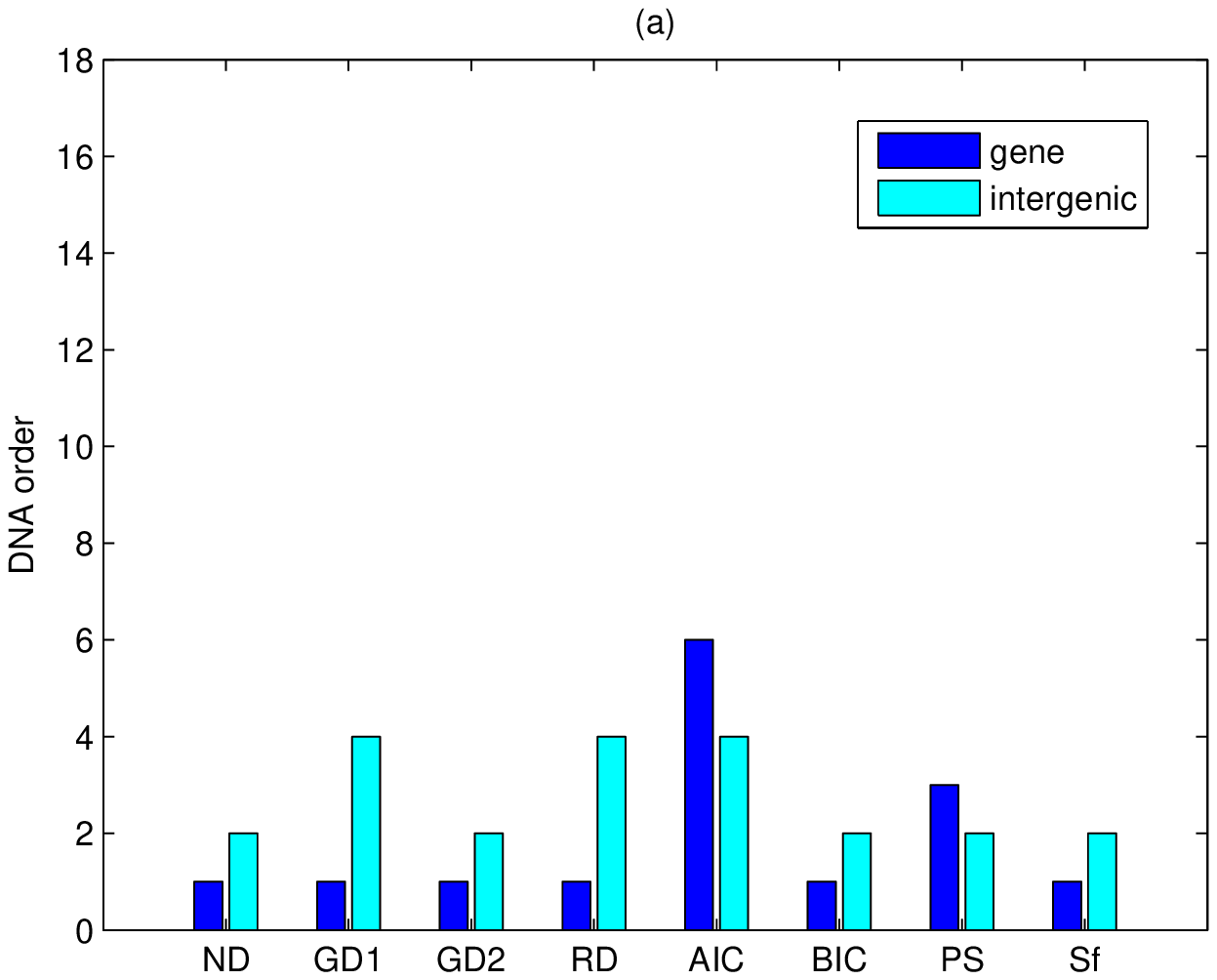}
\includegraphics[width=7cm]{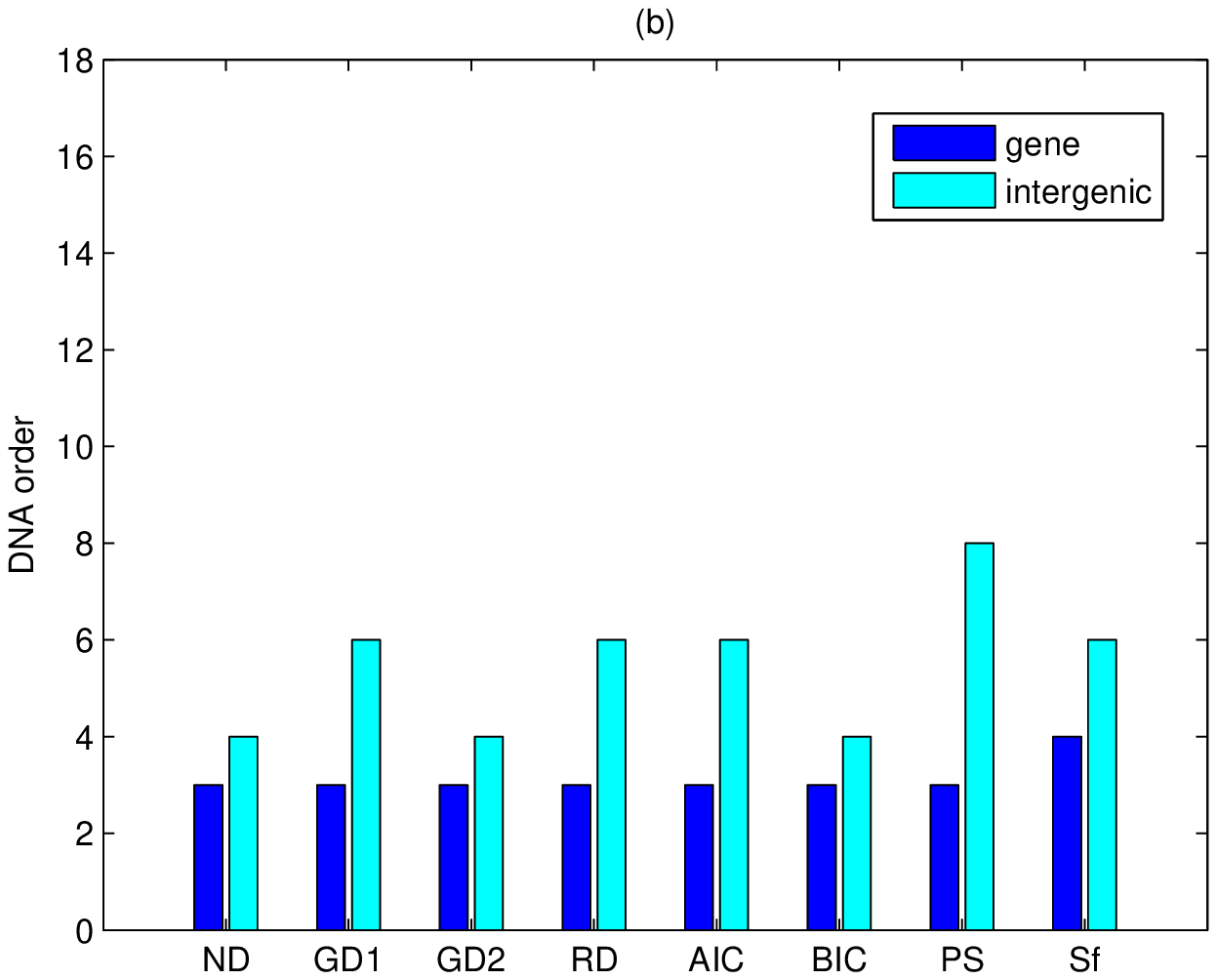}}}
\centerline{\hbox{\includegraphics[width=7cm]{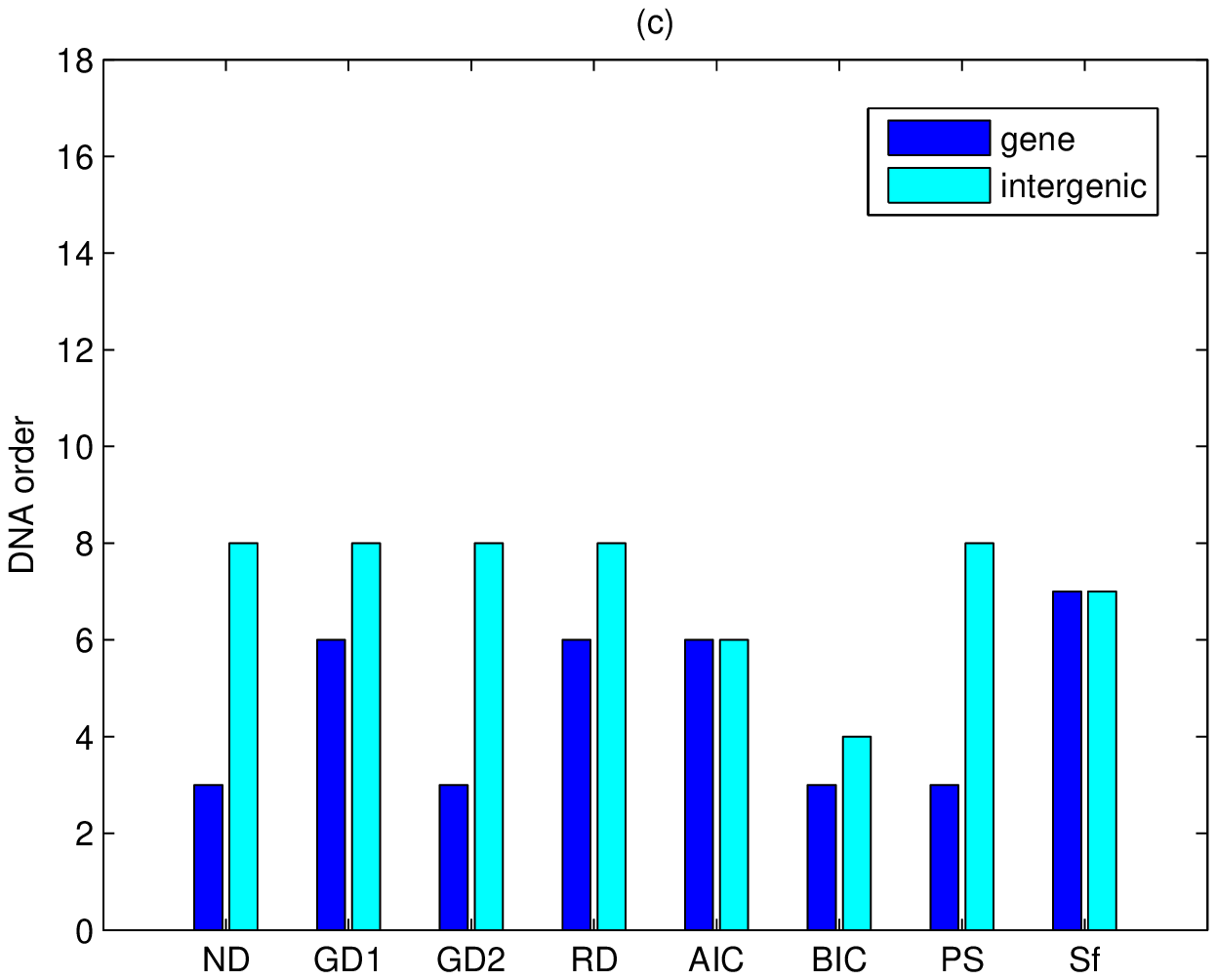}
\includegraphics[width=7cm]{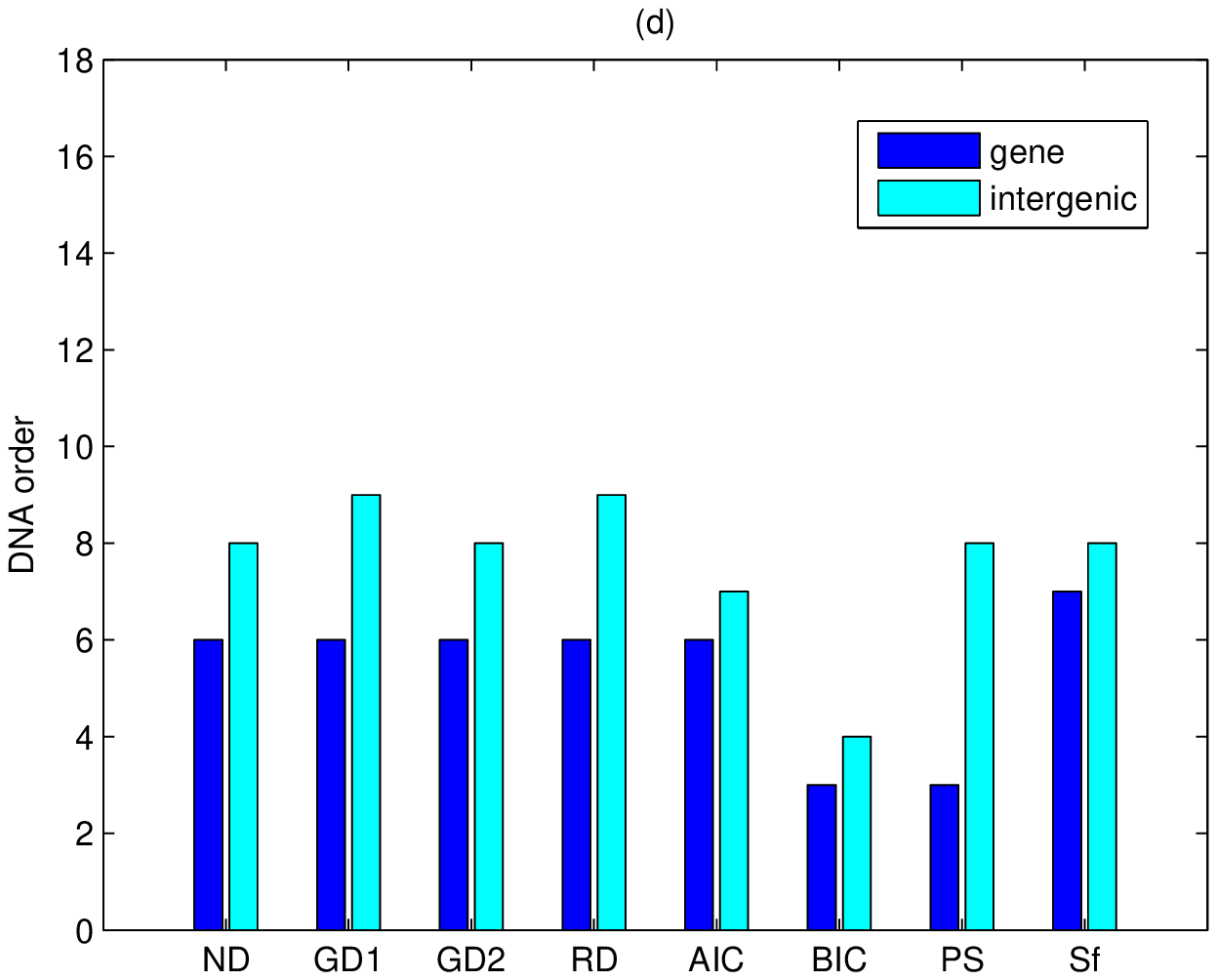}}}
\centerline{\includegraphics[width=7cm]{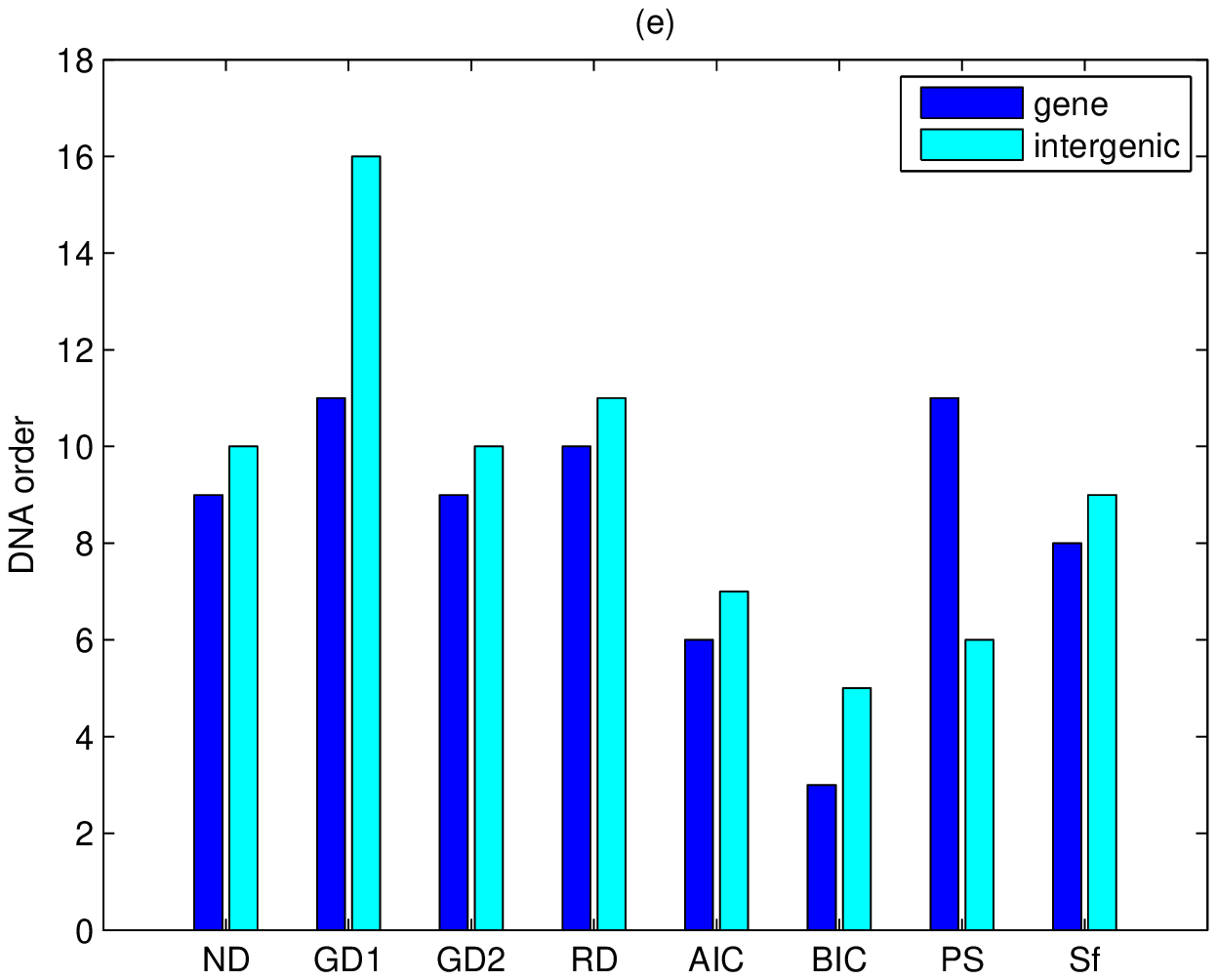}} \caption{The estimated order $L$ by ND, GD1, GD2, RD, AIC, BIC, PS and Sf of sequences of purine
and pyrimidine ($K=2$) from genes and intergenic regions of Chromosome 1 of the plant A\emph{rabidopsis} \emph{thaliana}, as indicated in the legend. The sequence lengths are
(a) $N=8000$, (b) $N=16000$, (c) $32000$, (d) $64000$ and (e) $128000$.} \label{fig:estimationorderDNA}
\end{figure}

We estimate the order $L$ of a hypothesized Markov Chain underlying Chromosome 1 of plant A\emph{rabidopsis} \emph{thaliana} by the three parametric
tests ND, GD1 and GD2, the RD, as well as the criteria of AIC, BIC, PS and Sf. The computations are done for both genes and intergenic regions of length $N=8000, 16000,
32000, 64000$ and $128000$ and for $K=2$ (purines, pyrimidines).
As shown in Figure~\ref{fig:estimationorderDNA}, the order estimated by any of the
four criteria based on CMI, and for both genes and intergenic regions, increases with the length $N$ of the DNA sequence, indicating the presence of a Markov chain of a
very large order (larger than the maximum order that can be detected for this $N$) or a chain with long range correlations.
The limits of detectable order for $N=8000$ (Figure~\ref{fig:estimationorderDNA}a) are $L=4$ for intergenic regions, obtained by GD1, RD and AIC, and
$L=6$ for genes,obtained by AIC whereas all four CMI-based criteria estimate $L=1$). The largest estimations of $L$ increase for $N=16000$ to $L=8$
and $L=4$ for intergenic regions and genes, respectively (Figure~\ref{fig:estimationorderDNA}b). The criterion of Sf gives about the same pattern of
increasing estimated order with $N$ and larger estimate of $L$ for intergenic regions than for genes. On the other hand, the estimated $L$ from the
criteria AIC, BIC and PS changes irregularly with $N$ and is not always larger for the intergenic regions, giving inconclusive results. The agreement
of $L$ estimation by GD1 and RD is remarkable, both giving exactly the same estimate for any of the two DNA types and for any but the largest length
$N=128000$. For $N=128000$ (Figure~\ref{fig:estimationorderDNA}e), the difference is small for genes with GD1 estimating $L=11$ and RD $L=10$, and
larger for intergenic regions with GD1 giving $L=16$ and RD $L=11$. The other two CMI based criteria, ND and GD2, give estimates of $L$ close to
these of GD1 and RD, and so does Sf but tending to give somewhat smaller estimate of $L$ as $N$ increases. The overall results suggest that the
symbol sequence of intergenic regions tend to have larger order and thus being more consistent to the hypothesis of long range correlation. This is
confirmed by the four CMI based criteria and Sf, but RD and GD1 in addition turn out to be able to estimate large $L$, as justified also by the
simulation results.


\section{Conclusions}
\label{sec:Conclusions} In this work we propose and assess parametric tests of significance of the conditional mutual information (CMI) for the
estimation of the order of Markov chain. The null distribution of CMI is approximated by the normal distribution and two different approximations of
gamma distribution. Simulations showed that among the three parametric tests the one based on gamma distribution (GD1) performed best for any Markov
chain order $L$ and number of symbols $K$ and even for short lengths of symbol sequences. The practical aim of the study was to investigate whether a
parametric test can reach the order estimation accuracy of the respective randomization test (RD), recently implemented and found to be compatible
and often better than the known order estimation criteria. The simulation study confirmed that GD1 performs similarly to RD and both compare
favorably to other known criteria (AIC, BIC, the Peres and Shields estimator and the criterion of Men\'{e}ndez et al. \citep{Mene06,Mene11}).

Having established the equivalence of performance of GD1 and RD,
the advantage of GD1 is the computational efficiency, allowing the
order estimation based on CMI to be possible for very long symbol
sequences, such as the DNA sequences. Obviously, RD applied with a
number $M$ of randomized sequences (in this work we used $M=1000$)
requires about $M$ times more computation time than GD1, and thus
application of RD is prohibitive for very long symbol sequences.
This was the case of DNA sequences, and for $N=128000$, RD was
running on a PC Intel Core CPU $2,83$GHz $3,5$GB RAM for about $2$
days.

Using the parametric and randomization tests, as well as the Sf criterion on purine and pyrimidine sequences of genes and intergenic regions from the
Chromosome 1 of plant A\emph{rabidopsis} \emph{thaliana}, we could establish an increase of the estimated order with the length of the DNA sequence,
indicating the presence of either a very large Markov chain order not reached by the tested sequence lengths or long range correlations (this is
further explored in a focused study in \cite{Papapetrou14}). Further, we  could also distinguish genes from intergenic regions as lower order was
estimated in genes, which consists of coding and non-coding parts, than in intergenic regions which contains non-coding parts exclusively.


\bibliographystyle{elsarticle-num}


\end{document}